\documentclass[structabstract]{aa}
\usepackage{natbib,ifthen,setspace,txfonts}
\usepackage{graphicx}
\usepackage{appendix}
\usepackage{lscape}
\begin{document}

\title{The mass-loss return from evolved stars to the Large Magellanic Cloud}
\subtitle{V. The GRAMS carbon-star model grid\thanks{The model grid is available at the CDS via anonymous ftp to {\tt cdsarc.u-strasbg.fr} (130.79.128.5) or via {\tt http://cdsweb.u-strasbg.fr/cgi-bin/qcat?J/A+A/}}}

\date{Received / Accepted}
\authorrunning{S. Srinivasan et al.}
\titlerunning{The GRAMS Carbon-Star Model Grid}

\author{S. Srinivasan\inst{\ref{inst1}}, B. A. Sargent\inst{\ref{inst2}} \& M. Meixner\inst{\ref{inst2}}}   

\institute{Institut d'Astrophysique de Paris, 98 bis Boulevard Arago, Paris 75014, France, email: {\tt srinivas@iap.fr}\label{inst1} \and Space Telescope Science Institute, 3700 San Martin Drive, Baltimore, MD 21218, USA\label{inst2}}
	        
\abstract
{Outflows from asymptotic giant branch (AGB) and red supergiant (RSG) stars inject dust into the interstellar medium. The total rate of dust return provides an important constraint to galactic chemical evolution models. However, this requires detailed radiative transfer (RT) modeling of individual stars, which becomes impractical for large data sets. An alternative approach is to select the best-fit spectral energy distribution (SED) from a grid of dust shell models, allowing for a faster determination of the luminosities and mass-loss rates for entire samples.}
{We have developed the {\em G}rid of {\em R}SG and {\em A}GB {\em M}odel{\em S} (GRAMS) to measure the mass-loss return from evolved stars. The models span the range of stellar, dust shell and grain properties relevant to evolved stars. The {GRAMS} model database will be made available to the scientific community. In this paper we present the carbon-rich AGB model grid and compare our results with photometry and spectra of Large Magellanic Cloud (LMC) carbon stars from the SAGE ({\em Surveying the Agents of Galaxy Evolution}) and SAGE-Spec programs.}
{We generate models for spherically symmetric dust shells using the {\bf 2D}ust\ code, with 
hydrostatic models for the central stars. The  model photospheres have effective temperatures between 2600 and 4000 K and luminosities from $\sim$2000 L$_\odot$\ to $\sim$40000 L$_\odot$. Assuming a constant expansion velocity, we explore five values of the inner radius $R_{\rm in}$ of the dust shell (1.5, 3, 4.5, 7 and 12 $R_{\rm star}$). We fix the outer radius at 1000 $R_{\rm in}$. Based on the results from our previous study, we use amorphous carbon dust mixed with 10\% silicon carbide by mass. The grain size distribution follows a power-law and an exponential falloff at large sizes. The models span twenty-six values of 11.3 $\mu$m\ optical depth, ranging from 0.001 to 4. For each model, {\bf 2D}ust\ calculates the output SED from 0.2 to 200 $\mu$m.} 
{Over 12\,000 models have dust temperatures below 1800 K. For these, we derive synthetic photometry in optical, near-infrared and mid-infrared filters for comparison with available data. We find good agreement with magnitudes and colors observed for LMC carbon-rich and extreme AGB star candidates from the SAGE survey, as well as spectroscopically confirmed carbon stars from the SAGE-Spec study. Our models reproduce the IRAC colors of most of the extreme AGB star candidates, consistent with the expectation that a majority of these enshrouded stars have carbon-rich dust. Finally, we fit the SEDs of some well-studied carbon stars and compare the resulting luminosities and mass-loss rates with those from previous studies.}
{}


\maketitle

\section{Introduction}
During the final stages of their evolution, low- and intermediate-mass stars (0.8 to 8 M$_\odot$) ascend the asymptotic giant branch (AGB). AGB stars are prominent members of stellar populations of ages $\sim$0.2--2 Gyr. Characterized by mass-loss rates of up to $10^{-4}$ M$_\odot$ yr$^{-1}$, stars in the AGB phase inject a significant fraction of their mass into the interstellar medium (ISM). This mass loss is thought to occur in two steps: stellar pulsations first levitate material to the cool, outer layers where dust grains form. Interaction with stellar photons then accelerates the dust grains, which in turn drag the gas along with them \citep[{\it e.g.},][]{GoldreichScoville1976,HofnerDorfi1997,Wachteretal2002,Hofner2009}. AGB stars of masses $\sim$1--4 M$_\odot$\ undergo the third dredge-up process \citep[{\it e.g.},][]{Iben1983,Karakasetal2002} which transports the products of nuclear reactions, including carbon, into the outer layers. Each dredge-up event increases the carbon abundance relative to oxygen until eventually the C/O ratio exceeds unity and carbon stars are born.
Carbon dust is more efficient at absorbing optical photons \citep[{\it e.g.},][]{WallersteinKnapp1998} and it has higher emissivity at infrared (IR) wavelengths. High rates of mass loss from C--rich AGB stars make them major contributors of atomic carbon and carbonaceous dust grains \citep{Dwek1998,Matsuuraetal2009,Srinivasanetal2009} to the ISM and may eventually be assimilated into star-forming regions. The thousands of carbon stars present in galaxies with intermediate-age stellar populations contribute substantially to their integrated bolometric and near-IR (NIR) luminosities \citep[][]{Frogeletal1990,Maraston1998}, and this contribution is somewhat higher at lower metallicity \citep[{\it e.g.}, Fig. 13 in ][]{Maraston2005}. Therefore, we must quantify the carbon-star dust output in order to study the dust cycle in galaxies as well as constrain stellar population synthesis models.

The study of Milky Way AGB stars is inhibited by the presence of substantial interstellar extinction, and large uncertainties in distance determinations. The Large Magellanic Cloud (LMC) offers the combination of proximity \citep[$\sim$50 kpc;][]{vanLeeuwenetal2007}, low line-of-sight extinction \citep[$E(B-V)\sim 0.075$ mag;][]{Schlegeletal1998} and favorable orientation \citep[$\sim$24$^\circ$;][]{WeinbergNikolaev2001}. These properties allow in-depth studies of the entire LMC AGB population. One such study, the {\it Spitzer} Space Telescope \citep{Werneretal2004} Legacy program SAGE \citep[Surveying the Agents of a Galaxy's Evolution;][]{Meixneretal2006}, imaged a 7$\times$7$^\circ$ area centered on the LMC and found over 6 million point sources, including thousands of carbon-star candidates \citep{Blumetal2006,Srinivasanetal2009}. Follow-up spectroscopy from the Infrared Spectrometer \citep[IRS,][]{Houcketal2004} on {\it Spitzer} was obtained as part of the SAGE-Spec program \citep{Kemperetal2010}. 

SAGE provides an ideal dataset for AGB studies. \citet{Srinivasanetal2009} (hereafter, Paper~I) calculated mid-IR (MIR) excess fluxes for AGB candidates identified from SAGE photometry and used these to estimate the total dust injection rate into the LMC. A more precise estimate for the injection rate requires detailed radiative transfer (RT) modeling of the spectral energy distribution (SED) of each star in the candidate list. Many authors have computed such detailed models for LMC carbon stars \citep[see, {\it e.g.},][]{vanLoonetal1999, Groenewegenetal2009, Srinivasanetal2010}. The computation of individual models becomes time-consuming for large samples such as the SAGE dataset. For this purpose, an alternative would be to compare the observed SEDs to those of pre-computed models. Ideally, such a grid of models should account for the photospheric absorption due to atomic and molecular species \citep[{\it e.g.},][]{Gautschy-Loidletal2004,Aringeretal2009} and dynamical effects such as pulsation-driven shocks \citep[{\it e.g.},][]{Hofneretal2003,Mattssonetal2007,Mattssonetal2010}, as well as the reprocessing of stellar radiation by dust \citep[{\it e.g.},][]{Groenewegen2006}. The effects of stellar evolution can be folded into the grid either by performing the above calculations on stars sampled along AGB evolutionary tracks \citep[see, {\it e.g.},][]{Mattssonetal2007} or by folding the effects of dust into stellar population modeling \citep[{\it e.g.},][]{Marigoetal2008,Gonzalezetal2010}.

The method of comparison with model grids is suitable for quickly constraining the general properties of large photometric datasets such as the LMC SAGE set. With the intention of quick SED fitting in mind, we have generated the {\em G}rid of {\em R}SG and {\em A}GB {\em M}odel{\em S} (GRAMS). The GRAMS\ grid consists of radiative transfer models computed using model stellar photosphere fluxes and assuming various values of stellar and dust shell parameters.
While some model grids are available in the literature \citep[{\it e.g.},][]{Groenewegen2006}, they typically use only a few template solar-metallicity photospheric spectra as the basis for the grid. Our models cover a large region of the parameter space spanned by AGB stars. One of the motivations for our study is the recent availability of large sets of photosphere models with an improved treatment of molecular spectral features of low-metallicity stars \citep{Gautschy-Loidletal2004,Aringeretal2009}. This enables us to probe the dependence of mass loss over a large range of stellar parameters. Other studies typically equate the dust temperature at the inner radius to the dust condensation temperature, which is usually fixed at 1000 K for carbon dust \citep[see, {\it e.g.},][]{vanLoonetal1999,Groenewegen2006}. We circumvent this assumption by adopting a more general treatment -- we specify the inner radius as input to the modeling and this automatically determines the temperature of the dust in the shell when the model is computed. The output SEDs are very sensitive to the inner radius $R_{\rm in}$ \citep[see, {\it e.g.},][]{Srinivasanetal2010} and we incorporate this dependence in our grid by computing models for different $R_{\rm in}$ values. We thus provide a large grid of models that is complementary to the currently available grids constructed by other authors. In previous papers in this series, we determined the properties of oxygen-rich \citep[][hereafter Paper~II]{Sargentetal2010a} as well as carbonaceous \citep[][hereafter Paper~III]{Srinivasanetal2010} dust grains for use with the grid. \citet{Groenewegen2006} incorporate different dust species in their grid. While the current incarnation of our grid is computed for a fixed set of dust properties, we will investigate other types of dust in future versions.

Fitting the SEDs of our entire dataset will allow us to investigate the LMC AGB mass-loss return. In this paper, we provide the details of the GRAMS\ carbon-star grid. Our O--rich grid is described in a companion paper  \citep[][hereafter Paper~IV]{Sargentetal2011}. The remainder of the paper is organized as follows. We detail the parameter selection for the carbon-star grid in Sect. \ref{sec:procedure} and describe the computational procedure and the calculation of synthetic photometry for the resulting grid in Sect. \ref{sec:computation}. In Sect. \ref{sec:results}, we compare the GRAMS\ synthetic photometry with SAGE observations on color--magnitude as well as color--color diagrams. As a further validation of the grid, we perform simple chi-squared fits to the SEDs of some sources that have been previously studied in detail. We present our conclusions in Sect. \ref{sec:summary}.

\section{The Model Grid}
\label{sec:procedure}
We use the {\bf 2D}ust\ code \citep{UetaMeixner2003} to populate the grid. {\bf 2D}ust is a radiative transfer code for axisymmetric systems. While the mass loss is spherically symmetric for most of the AGB phase, the highly evolved AGB stars as well as their post-AGB successors exhibit bipolar geometries \citep[{\it e.g.},][]{Uetaetal2000,Meixneretal2002}. The dust around post-AGB stars is also similar to AGB star circumstellar dust. We would like to produce models for these objects as well, but in this paper we assume spherically symmetric shells for simplicity. We will consider the effect of non-spherical geometries in future versions of the grid. In order to solve the radiative transfer equation, {\bf 2D}ust discretizes the dust shell into a 2-dimensional polar grid. The code then computes the radiation field at each grid point, discretizing the field into a set of incoming rays (``characteristics") that converge on the grid point from all directions. During each iteration, the code ensures self-consistency by requiring a global luminosity constancy throughout the dust shell. For a given stellar spectrum, dust shell geometry (inner and outer radius, density variation) and a set of dust grain properties (species, optical depth at a reference wavelength, grain size distribution), {\bf 2D}ust\ calculates the luminosity\footnote{Calculation of the luminosity requires the distance to the star, which we set at 50 kpc.} and mass-loss rate for the system. The code then solves the radiative transfer for the output star+dust spectrum. In this section, we discuss our parameter selection and provide details of our computational procedure. Table \ref{tab:params} lists some of the input/output parameters as well as the range in each parameter covered by the GRAMS\ carbon-star grid.

\begin{table*}
\caption{GRAMS\ carbon-star model grid parameter coverage}
\label{tab:params}
\centering
\begin{tabular}{ll}
\hline\hline\noalign{\smallskip}
\multicolumn{2}{l}{\bf Photosphere model\tablefootmark{a}}\\
\noalign{\smallskip}
$L_*$ (L$_\odot$) & $\sim$1100 to $\sim$26000\\
T$_{\rm eff}$ (K) & 2600 to 4000 (100)\tablefootmark{b}\\
$\log{g[{\rm cm~s^{-1}}]}$ & $-$1.0 to 0.0 (0.1)\tablefootmark{c}\\
M (M$_\odot$) & 1, 2, 3 and 5\\
C/O & 1.4, 2.0 and 5.0\\[0.25em]
\noalign{\smallskip}
\multicolumn{2}{l}{\bf Dust shell properties}\\
\noalign{\smallskip}
$R_{\rm in}$ ($R_*$) & 1.5, 3, 4.5, 7, 12\\
$R_{\rm{out}}$ ($R_{\rm in}$) & 1000\\
density profile & $\rho(r)\propto r^{-2}$\\
$v_{\rm{exp}} $(km~s$^{-1}$) & 10\\[0.25em]
\noalign{\smallskip}
\multicolumn{2}{l}{\bf Dust grain properties}\\
\noalign{\smallskip}
Species & AmC\tablefootmark{d}$+$SiC\tablefootmark{e}\\
SiC fraction & 10\%\\
$\tau$(11 $\mu$m) & $10^{-3}$ to $10^{-1}$ (5 per dex),\\
& 0.1 to 1 (0.1) and\\ 
& 1.5 to 4 (0.5)\\[0.25em]
Size distribution & KMH\tablefootmark{f}\\
 & $a_{\rm min}$($\mu$m) = 0.01\\
 & $a_{0}$($\mu$m) = 1\\
 & $\gamma$ = 3.5\\
\noalign{\smallskip}
\multicolumn{2}{l}{\bf Mass-loss rate and dust temperature}\\
\noalign{\smallskip}
$\dot{M}_{\rm dust}$ (M$_\odot$ yr$^{-1}$) & $1.5\times 10^{-12}$ to $2.1\times 10^{-7}$ \\[0.25em]
$\dot{M}_{\rm gas}$ (M$_\odot$ yr$^{-1}$)\tablefootmark{g} & $3.0\times 10^{-10}$ to $4.3\times 10^{-5}$ \\[0.25em]
$T_{\rm in}$ (K) & 710 to 1800\tablefootmark{h}\\
\hline\hline
\end{tabular}
\tablefoot{
\tablefoottext{a}{Photosphere model from \citet{Aringeretal2009} (A09).}
\tablefoottext{b}{Where applicable, parameter increments are supplied in parentheses.}
\tablefoottext{c}{No A09 model photospheres were available for $\log{g}=-0.1$}
\tablefoottext{d}{Amorphous carbon grains, $\rho=1.8$ g cm$^{-3}$, optical constants from \citet{Zubkoetal1996}.}
\tablefoottext{e}{SiC grains, $\rho=3.22$ g cm$^{-3}$, optical constants from \citet{Pegourie1988}.}
\tablefoottext{f}{Size distribution from \citet{Kimetal1994}:\\ $n(a)\sim a^{-\gamma}\exp{\left(-a/a_0\right)}$ with $a>a_{\rm min}$.}
\tablefoottext{g}{Assuming a gas:dust ratio of 200.}
\tablefoottext{h}{As explained in the text, the grid consists of models with temperatures cooler than 1800 K.}
}
\end{table*}

\subsection{Input for the grid}

\subsubsection{Stellar spectrum}
\label{subsec:photmodels}
We represent the central stars using the \citet{Aringeretal2009} COMARCS hydrostatic models of AGB star photospheres. Aringer et al. \nocite{Aringeretal2009} (hereafter, A09) take into account the contribution to the emergent spectrum from atomic and molecular absorption lines, which can cause the optical and near-infrared spectrum to deviate significantly from that expected from a blackbody at the same effective temperature. The A09 set consists of spherically symmetric COMARCS hydrostatic models calculated with opacities generated for CO, C$_2$, CN, C$_2$H$_2$, HCN, C$_3$ and a number of other molecules using the COMA (Copenhagen Opacities for Model Atmospheres) code. The parameter coverage of the models reflects the range of effective temperatures, surface gravities and C/O ratios predicted by synthetic evolution models. These are the most accurate models available at present. A09 find good agreement between their models and data for warmer AGB stars. They do not account for dynamical effects and the processing of starlight by dust. In addition to an entire grid calculated for solar metallicity, there are subsets of models for metallicities of $Z$=0.33 and 0.1 Z$_\odot$, corresponding to the Magellanic Clouds \citep[$Z_{\rm LMC}/Z_\odot$$\sim$0.3--0.5 and $Z_{\rm SMC}/Z_\odot$$\sim$0.1--0.2;][]{Dufouretal1982,Bernardetal2008}. As we are interested in first reproducing the observed range of colors for the LMC, we selected only their $Z=0.33$ Z$_\odot$\ subset of 131 models of masses $\leq 5$ M$_\odot$. We briefly outline the range of values covered by this subset. For details on the entire grid, we refer the reader to Sec 2.1 of A09. 

The effective temperatures of the $Z$=0.33 models (hereafter, the {\em LMC set}) range from 2600 K to 4000 K in increments of 100 K. While carbon stars may have lower effective temperatures than this range covers \citep{Mattssonetal2008}, the A09 models provide adequate coverage for the LMC C--rich sample \citep{Groenewegenetal2009}. As noted in A09, most models with temperatures hotter than 3500 K will not be on the AGB, but they are included in order to model carbon-rich objects in the post-AGB phase. The surface gravities range from $\log{(g[{\rm cm~s}^{-1}])}=0$ to $-1$ in steps of 0.1, but not all $\log{g}$ values are available for a given temperature. In general, the cooler the temperature, the lower the minimum $\log{g}$ available. The LMC set was calculated completely for a stellar mass of 2 M$_\odot$, resulting in 126 models. There are also three 1 M$_\odot$\ models with $T_{\rm eff}$=2600 K and 3000 K, as well as one $T_{\rm eff}$=3000 K model each for mass of 3 M$_\odot$\ and 5 M$_\odot$. A09 find that a change in stellar mass has only a minor effect on the near-IR color and bolometric magnitude (see their Fig. 8). These models have C/O ratios of 1.4, 2 and 5. Higher C/O ratios are expected at lower $Z$ due to the under-abundance of oxygen. Due to this fact, the C/O ratio can become significantly higher than unity within the first few dredge-up events. For the range of model parameters corresponding to each temperature range, see Table 1 in A09. The luminosities of the selected models are roughly in the range 1100 -- 26\,000 $L_\odot$ ($M_{\rm bol} = -2.9$ to $-6.3$).\\

In Fig. \ref{fig:lf}, we compare this range to the luminosity distribution calculated for SAGE C--rich and extreme AGB candidates. \citet{Marigoetal1999} estimate that the observed LMC carbon-star luminosity function extends between $M_{\rm bol} = -3$ and $-6.5$; these numbers are in good agreement with ours\footnote{We calculate luminosities higher than the classical AGB limit \citep[$M_{\rm bol} = -7.1$,][]{Paczynski1970} for a handful of our extreme AGB candidates; while a majority of these are well-studied O--rich AGB stars, there are no carbon stars in this subsample.}. IRAS~04496--6958 is the most luminous spectroscopically confirmed carbon star in the LMC \citep[$> 30\,000~L_\odot$; see, {\it e.g},][]{Specketal2006}. Based on SAGE photometry alone, we estimate its luminosity to be about 35\,000 $L_\odot$ ($M_{\rm bol} \approx -6.7$) which is still below the classical luminosity limit. Fig. \ref{fig:lf} also shows the GRAMS luminosity coverage. To fit the SEDs of the most luminous carbon stars, we may have to scale up the luminosities of our models in a procedure similar to that adopted in Sargent et al. (Paper~IV) for RSGs and O--rich AGB stars. Our aim in constructing the GRAMS model grid is to span the range of parameters observed for LMC carbon stars; it is possible that in this attempt we may generate some models with extreme luminosities that are not representative of real carbon stars and may never be matched to data. We do not use luminosity scaling in the current version of the carbon star grid; however the scaling is easy to apply once the model grid has been generated, and we will consider such a scheme when we perform SED fitting of the SAGE data in upcoming papers.

\subsubsection{Shell geometry}
We assume a constant mass-loss rate at a constant outflow velocity $\varv_{\rm exp}$, leading to an inverse-square density distribution in the shell. Models of stationary winds \citep[{\it e.g.},][]{Woitke2006} show that the wind is accelerated from rest near the stellar surface to an almost constant speed within a few stellar radii. The acceleration of the wind therefore affects the density profile in the regions of the dust shell that contribute significantly to the optical extinction and mid-IR emission. In the current paper, we do not take this radial dependence of the outflow velocity into account; ignoring such complications enables us to generate a full grid spanning the observed range for a small set of parameters while allowing us to compensate for the simplifications at a later time. We also ignore any dependence of the expansion velocity on the metallicity or gas:dust ratio. As the dust mass-loss rate calculated by the code is directly proportionate to the value of $\varv_{\rm exp}$\ chosen (see, {\it e.g.}, Eq. 2 in Paper~III), it is straightforward to incorporate the effect of changing $\varv_{\rm exp}$\ once the grid has been populated. Following our discussion in Paper~III, we use $\varv_{\rm exp}$\ = 10 km~s$^{-1}$\ for all our models. {\bf 2D}ust accepts a user-defined density function, which can be used to study time-dependent mass loss. We will also consider this possibility in future versions of the grid.

As discussed in Paper~III, the mass-loss rate and output spectrum are very sensitive to the value of the inner radius $R_{\rm in}$. The inner radius determines the hottest temperature of the dust, which is one of the {\bf 2D}ust\ output parameters. We calculate models for $R_{\rm in}$ = 1.5, 3, 4.5, 7 and 12 stellar radii. Simple energy-balance estimates suggest that amorphous carbon dust should form within a few stellar radii \citep{Hofner2007}. A lower limit of about 1.3 $R_{\rm in}$\ is suggested for SiC formation in extreme carbon stars by \citet{Specketal2009}. The range considered here also agrees with observations of Galactic carbon stars \citep[{\it e.g.}, IRC+10216;][]{Danchietal1995} and results from RT modeling of LMC stars \citep[{\it e.g.},][]{vanLoonetal1999}. We note here that {\bf 2D}ust\ uses the inner radius as input to determine the dust temperature ($T_{\rm in}$) as a function of radius in the shell. This means that we cannot directly restrict the range of $T_{\rm in}$\ before the grid is computed. However, we filter out models with high and/or unphysical dust temperatures from the grid once it is generated (see Sect. \ref{subsec:filtering} for details).

The outer radius determines the total amount of mass in the shell and the mass-loss timescale. While these are important quantities, we focus on obtaining mass-loss rates, which are only weakly sensitive to changes in the value of the outer radius. We ensure that the outer radius is large enough so that we do not miss any contributions to the flux from the outermost regions. While modeling the shell around OGLE LMC LPV 28579 in Paper~III we found that a value of $R_{\rm out}$\ = 1000 $R_{\rm in}$\ was sufficiently large to satisfy this condition. We use the same value for all the models in our grid. For $\varv_{\rm exp}$ = 10 km~s$^{-1}$ and $R_{\rm out}$\ = 1000 $R_{\rm in}$, the mass-loss timescales corresponding to the smallest and largest circumstellar envelopes in our grid are 548 yr and 19\, 620 yr respectively.

\subsubsection{Dust grain properties}
\label{subsec:dgprop}
We modeled the SAGE photometry and SAGE-Spec spectrum of OGLE LMC LPV 28579 in Paper~III for the purpose of selecting a set of carbonaceous dust grain properties for GRAMS. As a result, we choose a mixture of amorphous carbon \citep[AmC; optical constants from][]{Zubkoetal1996} with 10\% silicon carbide by mass \citep[optical constants from][]{Pegourie1988}. As input, {\bf 2D}ust\ also requires the optical depth specified at a reference wavelength. In Paper~III, we used the SiC feature observed in the SAGE-Spec spectrum of LPV 28579 to constrain the optical depth as well as the SiC content of our dust model, so it was convenient to specify the optical depth at 11.3 $\mu$m. We follow the same practice in this paper, because the dust composition used in the grid is identical to that of Paper~III.\footnote{For reference, the optical depth at 1 $\mu$m is about 10.64 times its value at 11.3 $\mu$m for the chosen set of dust properties.} We will revise the current convention when we incorporate more dust types into our grid. The values of $\tau_{11.3}$ in our grid range from $10^{-3}$ to 4. We consider five optical depths per decade between $10^{-3}$ and 0.1 (1, 2, 4, 6 and 8 $\times 10^{-3}$ and so on), and ten values between 0.1 and 1. Additionally, we calculate models for $\tau_{11.3}$ = 1.5, 2, 2.5, 3, 3.5 and 4, bringing the total to twenty-six unique optical depth values. The grain sizes are distributed according to $n(a)\propto a^{-\gamma}e^{-a/a_0}$ \citep[KMH distribution,][]{Kimetal1994} with power-law index $\gamma = 3.5$, minimum grain size $a_{\rm min} = 0.01$ $\mu$m\ and exponential scale height $a_0 = 1$ $\mu$m. For the AmC:SiC mixture considered here, the average grain size\footnote{For the GRAMS\ grid, averages over grain size space are computed by weighting according to grain surface area \citep[Harrington averaging scheme,][]{Harringtonetal1988}} is about 0.1 $\mu$m, the value typically used in single-size models \citep[{\it e.g.},][]{Groenewegen2006}. The \citet{Mathisetal1977} (MRN) distribution, given by $n(a) \propto a^{-\gamma}$ for $a_{\rm min} < a < a_{\rm max}$ places a strict limit on the maximum grain size. In this sense, the KMH model represents a more realistic grain size prescription. In Paper~III, we showed that the output spectrum showed only a weak dependence on the exponential factor $a_0$ of the KMH model beyond $a_0 \approx 0.1 \mu$m. Our choice of the KMH model thus means there is one less parameter to constrain. Owing to its maximum grain size requirement, the MRN distribution has a smaller average grain size. This results in higher absorption and therefore lower optical flux and more mid-IR emission. This effect was discussed in \citet{UetaMeixner2003}. In our study, we computed {\bf 2D}ust models for a few GRAMS models using the MRN prescription and found that the maximum difference in fluxes is less than 5\%.

\section{Generating The Grid}
\label{sec:computation}
\subsection{Preparation}
\label{subsec:preparation}
\begin{figure*}[!htb]
\center{\includegraphics[width=12cm]{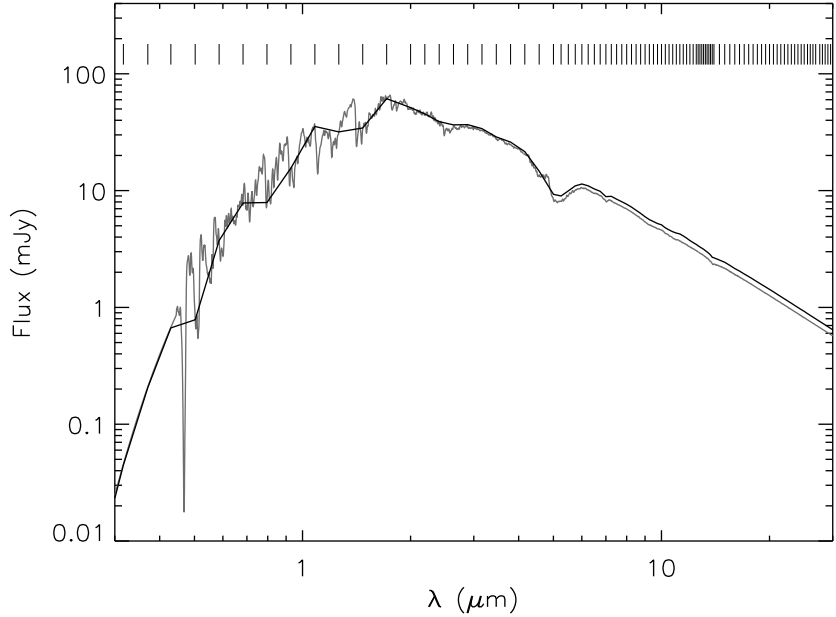}}
\caption[]{\citet{Aringeretal2009} photosphere spectrum (gray) with $T_{\rm eff}=3300$ K and $\log{(g[{\rm cm~s}^{-1}])}=-0.2$, showing many strong features at shorter wavelengths. The vertical dashes show the wavelengths at which this spectrum was sampled for input to {\bf 2D}ust. The resulting GRAMS output model with optical depth $\tau_{11.3}=2\times 10^{-3}$ (black) incorporates some of the variation in the optical.
\label{fig:spectralsample}}
\end{figure*}

We use 131 A09 photospheres and we explore five values for the inner radius as well as twenty-six optical depths, making a total of 17\,030 output models. Generating such a huge grid of models requires that {\bf 2D}ust\ be run in the non-interactive mode, with all the required information provided in the form of various input files (For details of the input format, we refer the user to Sect. 3.1 of the {\bf 2D}ust\ manual.\footnote{The {\bf 2D}ust\ manual is available at {\tt http://www.stsci.edu/science/2dust/2dust\_manual.pdf.gz}}) 

The A09 photospheric spectra comprise of fluxes at about 8\,000 wavelengths ranging from $\sim$0.44--25 $\mu$m. As we are interested in providing synthetic photometry over all the broadband filters available to us (optical U through MIPS24), we extrapolate the A09 fluxes onto a larger wavelength grid. On the long-wavelength side, we extend the grid beyond the MIPS24 band, to allow for future comparisons with MIPS 70 and 160 $\mu$m\ fluxes or other long-wavelength data, if available. However, this raw extrapolated spectrum cannot directly be fed into {\bf 2D}ust; the computational speed of each run is approximately linearly proportional to the number of points in the wavelength grid (Sect. 3.1.5 of the {\bf 2D}ust\ manual). We therefore sample the extrapolated spectrum at about 130 wavelengths ranging from 0.2 $\mu$m to about 200 $\mu$m. We sample more near- and mid-infrared wavelength points as we desire to fit the mid-IR photometry and spectra available as part of the SAGE and SAGE-Spec programs; the constraint on the total number of wavelength points then reduces the number of samples for $\lambda < 1~\mu$m. As a result of this undersampling, we expect that we may not reproduce many of the sharp atomic/molecular features observed in the optical spectra of AGB photospheres, which would affect the heating of dust grains as well as the broadband optical colors of the models with optically thin dust shells. This latter issue is of less concern since we are interested in reproducing the near- and mid-IR photometry. However, in order to have a more accurate treatment of dust heating, we will incorporate a better sampling in the optical in future versions of the grid. Fig. \ref{fig:spectralsample} demonstrates the effect of undersampling on the output model for a $T_{\rm eff}=3300$ K, $\log{(g[{\rm cm~s}^{-1}])}=-0.2$ photosphere with an optically thin ($\tau_{11.3}=2\times 10^{-3}$) dust shell. While the resulting GRAMS model does not reproduce all the narrow optical features, it shows some of the large-scale variation in the spectrum. The sampling described above also changes the integrated flux calculated for each model by at most a few percent -- we found that the luminosities calculated from the raw and sampled spectra agreed to within 5\% for 115 of the 131 A09 photospheres, with the maximum discrepancy being around 8\%. For consistency, we only use the post-sampled luminosities to tag our resulting models.

The number of discrete radial zones in the shell (NRAD) directly determines the total memory allocated to the code. Based on our tests, we chose to have 64 radial zones for all of our models. The code converges to a solution when, at each grid point, the fractional change in the integrated flux between consecutive iterations is less than a user-defined tolerance level. This level is specified in the code in the form of the CONDITION parameter, which we fixed at 5$\times$10$^{-4}$ for all the models. The execution time is then most sensitive to the optical depth. There is a trade-off between output precision and execution efficiency, which are regulated internally by two parameters: one (VSPACE) sets the smallest line integration step size in terms of the local mean free path length, while the other (MXSTEP) sets the maximum number of allowed steps for line integration along each characteristic. The smallest step size is inversely proportionate to VSPACE value chosen, so large VSPACE values are required for higher precision in the results. However, VSPACE determines the actual number of steps along the characteristic, which cannot exceed MXSTEP. This latter value controls the array length for long characteristics and therefore the memory access time upon execution of the code. Therefore, we need to optimize the choices for MXSTEP and VSPACE as a function of optical depth in order for the code to converge in a reasonable amount of time. For this purpose, we ran a number of test models on a MacBook Pro laptop with a 3.06 GHz Intel Core 2 Duo processor and 8 GB RAM. The test models sampled optical depths over the entire range of values considered for the grid. These models explored various (MXSTEP, VSPACE) pairs, enabling us to select ideal values over different optical depth ranges. This information is incorporated into the input files for each model in the grid, and this input enables us to run {\bf 2D}ust\ in the non-interactive mode in an automated fashion.

{\bf 2D}ust tracks the temperature of the dust grains during each iteration. Assuming radiative equilibrium, the code uses these temperatures to obtain the corresponding intensities $\kappa_\nu B_\nu$ from a look-up table, which is computed on execution. This table consists of a temperature grid whose lower and upper limits are TBOT = 2.7 K and TTOP = 1000 K (the typical value chosen for the condensation temperature of carbon dust) by default. It is not easy, however, to constrain $T_{\rm cond}$ as it depends on many factors such as the C/O ratio and the gas pressure \citep[{\it e.g.}, Sect. 3.4 of][ and references therein]{Specketal2009}. A detailed treatment of $T_{\rm cond}$ is beyond the scope of our work. \citet{Specketal2009} suggest that carbon dust can form at temperatures above 1400 K even at low mass-loss rates. They also find that for high C/O ratios, graphite grains can form at temperatures of $\sim$1800 K. While our current models only consider amorphous carbon dust grains, we are interested in as few constraints on the output parameters as possible so as to be able to extend our treatment by including more grain types in the future if required. Therefore, we assume an upper limit of 1800 K for $T_{\rm cond}$. We allow for slightly higher dust temperatures in the look-up table by setting TTOP to 2000 K to avoid convergence issues (see Sect. \ref{subsec:filtering}), and we enforce the 1800 K constraint on $T_{\rm cond}$ once the grid is generated.

\subsection{Batch job submission}
We automated the {\bf 2D}ust\ execution using the Magique supercomputing cluster at the Institut d'Astrophysique de Paris. Magique is a cluster of 96 AMD operton nodes interconnected with a Gigabit ethernet network, 84 of which are reserved for scientific computation. Of these, we used the bisockets single and double core nodes (18 and 64 nodes respectively), which allowed 8 GB RAM per job. {\bf 2D}ust\ was compiled using the Intel FORTRAN compiler {\tt ifort} available on Magique. The batch queue system on the cluster is managed by {\tt PBSpro} v8.0, which automatically allocates a position in the queue for submitted jobs and continuously monitors their status. Our jobs were grouped according to the A09 photosphere used and they were submitted for execution in batches consisting of up to five sets of photospheres ({\it i.e.}, upto 650 models), depending on the queue status and the load on the cluster from other users. Depending on the optical depth, the time taken for a single model to converge ranged from under a minute to about two hours. We were able to generate the entire grid in under two weeks.

\subsection{Filtering out models based on dust temperature}
\label{subsec:filtering} 
Not all combinations of input parameters detailed in Tab. \ref{tab:params} will result in realistic or even physical representations of the dust around carbon stars. For instance, we expect that the output dust temperatures for high $T_{\rm eff}$, low $R_{\rm in}$\ and high $\tau_{11.3}$ may be higher than the dust condensation temperature $T_{\rm cond}$. If the local radiation intensity is high enough to warm the dust grains to temperatures beyond TTOP (the maximum value in the lookup table, which was set to 2000 K as described in Sect. \ref{subsec:preparation}), the code is unable to track these temperatures and intensities, resulting in unphysical (extremely large or negative) dust temperatures in the inner regions of the shell. We found that 1860 models did not converge because of this issue. As expected, these models consisted mainly of shells that were too close to the star ($R_{\rm in} = 1.5 R_{\rm star}$), but also included some models with larger $R_{\rm in}$ and $T_{\rm eff}$ values warmer than 3500 K. The number of {\em converged} models in our grid is therefore 15\,170. We then impose the constraint that the dust temperatures be cooler than the chosen condensation temperature ($T_{\rm cond} = 1800$ K, see Sect. \ref{subsec:preparation}), which eliminates 2915 more models. Our final grid thus consists of 12\,243 models with dust temperatures cooler than 1800 K. The number will reduce further if we adopt a lower $T_{\rm cond}$ value; for instance, there are only $\sim$9000 models with $T_{\rm in}$ $<$ 1400 K.

\subsection{Publicly available synthetic photometry}
For comparison with photometric data, we convolved the SED output from each {\bf 2D}ust\ model with the relative spectral response (RSR) curves for the various broadband filters. In particular, we produced synthetic photometry in this fashion for the MCPS UBVI
\footnote{The MCPS magnitudes were placed on the Johnson-Kron-Cousins UBVI system. The detector quantum efficiency (QE) curve available on the Las Campanas Observatory website at {\tt http://www.lco.cl/lco/telescopes-information/irenee-du\\-pont/instruments/specs/du-pont-telescope-direct-ccd-\\camera-ccd} was extrapolated to 0.3 $\mu$m. The QE was also assumed to drop linearly to zero from its value at 0.84 $\mu$m, the last wavelength provided on the graph, to 1.13 $\mu$m, the wavelength corresponding to a photon energy of 1.1 eV, which is the energy of the band gap of silicon. The B- and V-band transmission profiles were obtained from {\tt http://www.lco.cl/lco/telescopes-information/irenee-du\\-pont/instruments/website/direct-ccd-manuals/direct-ccd\\-manuals/3x3-filters-for-ccd-imaging} (Harris B profile LC-3013 and Harris V profile LC-3009 respectively). The U- and I-band filter profiles were obtained from I. Thompson (2009, private communication).}, 2MASS JHK$_{\rm s}$\footnote{The 2MASS filter relative spectral responses (RSRs) derived by \citet{Cohenetal2003} were obtained from the {\it 2MASS All-Sky Data Release Explanatory Supplement}, available at {\tt http://www.ipac.caltech.edu/2mass/releases/allsky/doc/\\
sec6\_4a.html.}} 
and {\it Spitzer} IRAC\footnote{RSRs available at {\tt http://ssc.spitzer.caltech.edu/irac/\\
calibrationfiles/spectralresponse}} 
and MIPS24\footnote{RSR available at {\tt http://ssc.spitzer.caltech.edu/mips/\\
calibrationfiles/spectralresponse/}} 
bands. We compare the synthetic photometry against SAGE data in the next section.  

Synthetic photometry was also derived for the AKARI \citep{Murakamietal2007} and WISE \citep{Wright2009} passbands\footnote{Filter response curves available at\\ {\tt http://www.ir.isas.jaxa.jp/ASTRO-F/Observation/RSRF/\\IRC\_FAD/index.html} and\\ {\tt http://www.astro.ucla.edu/$\sim$wright/WISE/passbands.html} respectively}. The AKARI S11 band (centered around 11 $\mu$m) as well as the WISE W3 (centered around 12 $\mu$m) band directly sample the 11.3 $\mu$m\ SiC feature in carbon stars and the $\sim$ 10 $\mu$m\ silicate feature in O--rich AGB stars. \citet{Itaetal2008} presented the AKARI survey of the LMC and discussed color--color and color--magnitude diagrams. The increase in SiC feature strength, therefore, is reflected in the [S11]--[L15] color. 

Spectra and synthetic photometry for the GRAMS\ O--rich and C--rich models will soon be available on the {\bf 2D}ust website at the Space Telescope Science Institute\footnote{\tt http://www.stsci.edu/science/2dust/grams\_models.cgi}. The format of the publicly-available data is still undergoing testing and will be refined through input from the end-users. In addition to the filters mentioned above, we can make synthetic photometry available for other filters on request.

\section{Results}
\label{sec:results}
In this section, we demonstrate the coverage of various parameters by the GRAMS\ grid. We discuss the luminosities and dust mass-loss rates. We also compare the GRAMS\ synthetic photometry with data for the SAGE C--rich and extreme AGB candidates as well as SAGE-Spec carbon stars on color--magnitude diagrams (CMDs) and color--color diagrams (CCDs). 

\begin{figure*}[!hbt]
\center{\includegraphics[width=12cm]{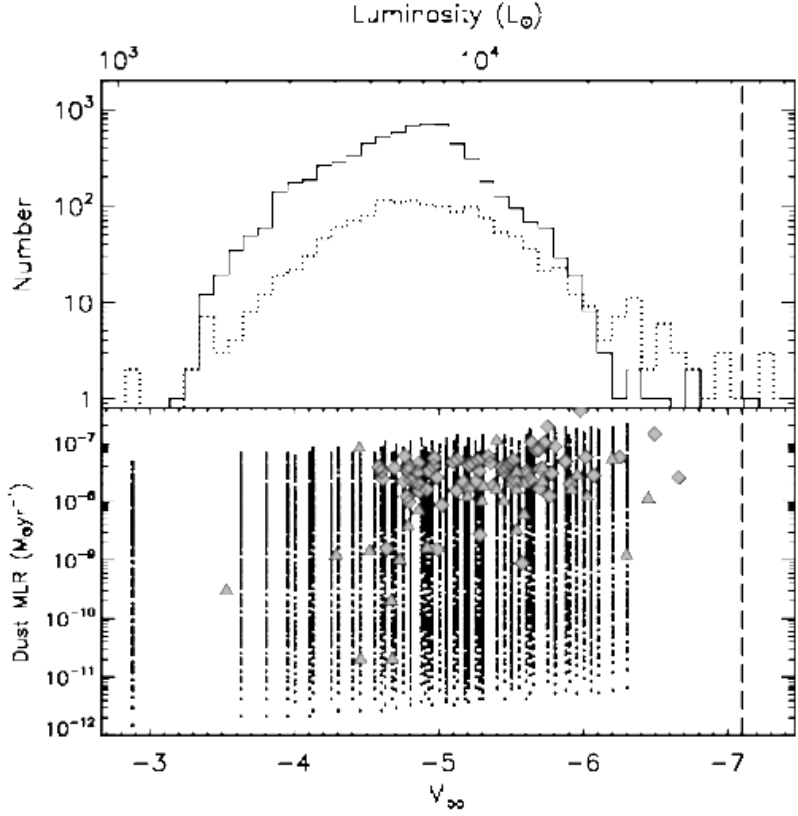}}
\caption[]{{\em Top}: The distribution of luminosities of LMC C--rich (solid) and extreme (dotted) AGB candidates from SAGE. The classical AGB luminosity limit is also shown (dashed line). {\em Bottom}: The range of luminosities and dust mass-loss rates covered by the GRAMS\ carbon star models. As discussed in Sect. \ref{subsec:filtering}, only models with $T_{\rm in}<1800$ K are shown. Gray symbols show the luminosities and mass-loss rates of LMC carbon stars as calculated from detailed modeling by \citet{vanLoonetal1999} (triangles) and \citet{Groenewegenetal2009} (diamonds).
\label{fig:lf}}
\end{figure*}

\subsection{Luminosities and dust mass-loss rates}
\label{subsec:results:lummlrcoverage}
The large number of A09 models provide us with a relatively dense coverage of the luminosities expected for LMC AGB stars. This fact is evident from Fig. \ref{fig:lf}, which compares the GRAMS\ model luminosities with the distribution of luminosities calculated in Paper~I from SAGE photometry for C--rich and extreme AGB candidates\footnote{The luminosities in Paper~I (Fig. 4), obtained using the trapezoidal rule, were overestimated for the brightest stars because of erroneous coefficients in the sum. In the present paper we have used the correct coefficients, leading to lower luminosities as shown in Fig. \ref{fig:lf}.}. The GRAMS\ models provide excellent coverage over the entire range of luminosities, except at the faint and bright ends. There is a gap visible between the two lowest available luminosities of 1100 L$_\odot$ and 2200 L$_\odot$. There are 50 candidates (8 C--rich, 42 extreme) brighter than the brightest C--rich GRAMS\ model available, and 60 candidates (45 C--rich, 15 extreme) fainter than $\sim$ 2200 L$_\odot$. 

The dust mass-loss rates in the GRAMS\ grid range from $1.5\times 10^{-12}$ M$_\odot$ yr$^{-1}$\ to $2.1\times 10^{-7}$ M$_\odot$ yr$^{-1}$. As mentioned in Sect. \ref{subsec:dgprop}, {\bf 2D}ust\ can only constrain the ratio $\dot{M}_{\rm dust}/$$\varv_{\rm exp}$, so the range of mass-loss rates produced by the grid depends on the value of $\varv_{\rm exp}$\ chosen. The lowest values of mass-loss rates calculated from the IR excesses for SAGE carbon star candidates in Paper~I were around $10^{-11}$ M$_\odot$ yr$^{-1}$. As we are interested in covering as much of the parameter space as possible, our grid contains mass-loss rates well below this limit. Carbon stars can have dust mass-loss rates up to a few times $10^{-7}$ M$_\odot$ yr$^{-1}$\ \citep[{\it e.g.},][]{Groenewegenetal2009}. Our grid does not currently have adequate coverage of the highest luminosities expected for carbon stars (see Sect. \ref{subsec:photmodels}), but the models can be scaled to these higher luminosities as in Paper~IV. This would also result in higher mass-loss rates. As an example, if we scale the brightest model ($L=2.6\times 10^{4}$ L$_\odot$) to the classical AGB limit ($L=5.3\times 10^4$ L$_\odot$) while keeping all the other parameters constant, the corresponding highest mass-loss rate is $3\times 10^{-7}$ M$_\odot$ yr$^{-1}$. The total (gas+dust) rate depends on the gas:dust ratio, $\Psi$. As AGB stars produce their own carbon, it is easier at lower metallicities to produce an excess abundance of carbon relative to oxygen; however, whether the gas:dust ratio shows a dependence on metallicity remains an open question \citep[see, {\it e.g.},][and references therein]{vanLoonetal2008}. Using $\Psi=200$, the value determined for Galactic carbon stars \citep{Jura1986}, the range of gas mass-loss rates covered by the grid is $3.0\times 10^{-10}$ M$_\odot$ yr$^{-1}$\ to $4.3\times 10^{-5}$ M$_\odot$ yr$^{-1}$. We plot the GRAMS\ dust mass-loss rates against their luminosities in Fig. \ref{fig:lf}. The luminosities and mass-loss rates determined for some LMC carbon stars that were modeled by \citet{vanLoonetal1999} and \citet{Groenewegenetal2009} are also shown for comparison. The GRAMS\ grid has good coverage of the range of luminosities and mass-loss rates calculated by these studies, except at very high luminosities and mass-loss rates -- in particular, there are three stars modeled by \citet{Groenewegenetal2009} with higher mass-loss rates. We would like to point out that this is no longer an issue if we adopt the same amorphous carbon dust opacities as in the Groenewegen et al. study (see Sect. \ref{subsubsec:G09compare} for details). The current optical depth limit of $\tau_{11.3} = 4$ was chosen based on the reddest sources in the SAGE study -- with the current set of models, we are able to produce redder mid-IR colors than those observed in the SAGE sample (see Sect. \ref{subsec:CMDCCD}). The choice was also based on results from our empirical study (Paper~I). As models with higher optical depth are computationally intensive, we decided to add optically thicker models in the future if required.

\subsection{Color--magnitude and color--color diagrams}
\label{subsec:CMDCCD}
The figures in this section compare the GRAMS\ synthetic photometry with SAGE O--rich, C--rich and extreme AGB candidates. These plots demonstrate that the GRAMS\ grid is able to reproduce the range of observed colors for carbon star candidates. We also include sources from the SAGE-Spec study that display molecular features and/or dust signatures typical of carbon stars \citep{Kemperetal2010,Woodsetal2010}. Only the subset of SAGE-Spec sources that had a full 5--37 $\mu$m\ spectrum are shown here. We refer the reader to \citet{Blumetal2006} for a detailed description of the various stellar populations observed in these diagrams and to Paper~IV for discussion on the locations of the oxygen-rich AGB candidates and models.

We would like to emphasize that comparisons on CMDs and CCDs are only shown in order to demonstrate the coverage of the observed colors and magnitudes, and we are careful not to over-interpret the agreement. Some of our models may be unrealistic, yet others may correspond to unphysical parameter values. While we have made an effort to avoid unrealistic models by filtering out those with very high dust temperatures, our aim was to span a large parameter space and good agreement with data does not necessarily validate a model in the grid. Similarly, good agreement with data on a color--color diagram does not necessarily imply that a model is a good fit to the data. Two models with significantly different values for the same parameter ({\it e.g.}, luminosity) but similar SED shapes might also end up close to each other on a color--color diagram, resulting in a large uncertainty for the parameter calculated for data based on just this diagram.

\subsubsection{K$_{\rm s}$\ {\it vs.} J--K$_{\rm s}$\ CMD}
\begin{figure*}[!htb]
\center{\includegraphics[width=12cm]{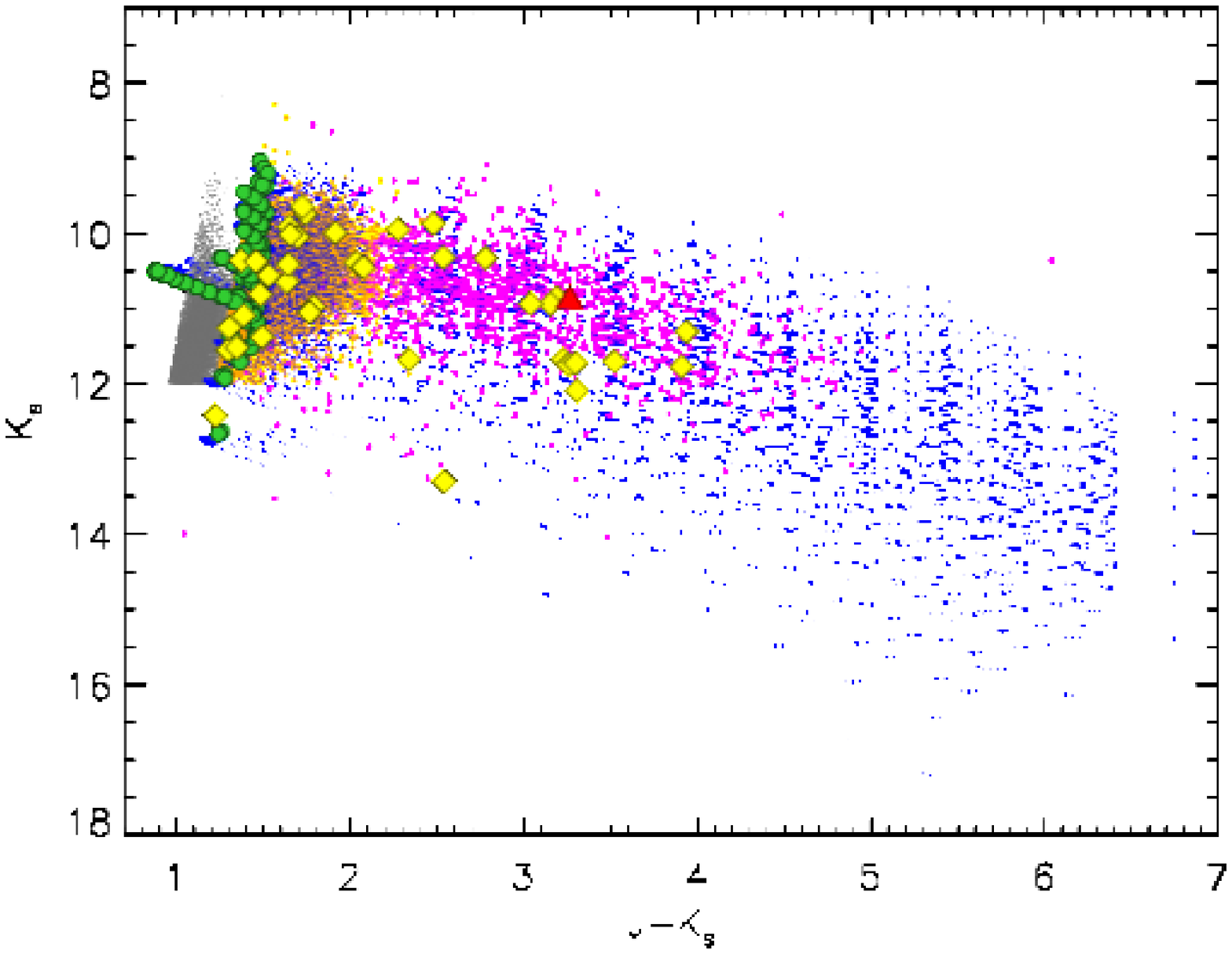}}
\caption[]{The K$_{\rm s}$\ {\it vs.} J--K$_{\rm s}$\ CMD shows the model grid (light blue points) overlaid onto the SAGE AGB candidates from Paper~I (gray: O--rich, orange: C--rich and pink: extreme). The yellow diamonds are SAGE sources identified as carbon-rich from their SAGE-Spec spectra. The large red triangle is OGLE LMC LPV 28579. Also shown are the \citet{Aringeretal2009} photospheres (green squares) used to generate the grid. The ``spike" running across the O--rich AGB candidates consists of models with $T_{\rm eff}$\ warmer than 3500 K. The cloud of models with $\tau_{11.3}=0.3$ is seen at J--K$_{\rm s}$\ $\sim$ 3 mag.\label{fig:jkcmd}}
\end{figure*}

\begin{figure*}[!htb]
\center{\includegraphics[width=12cm]{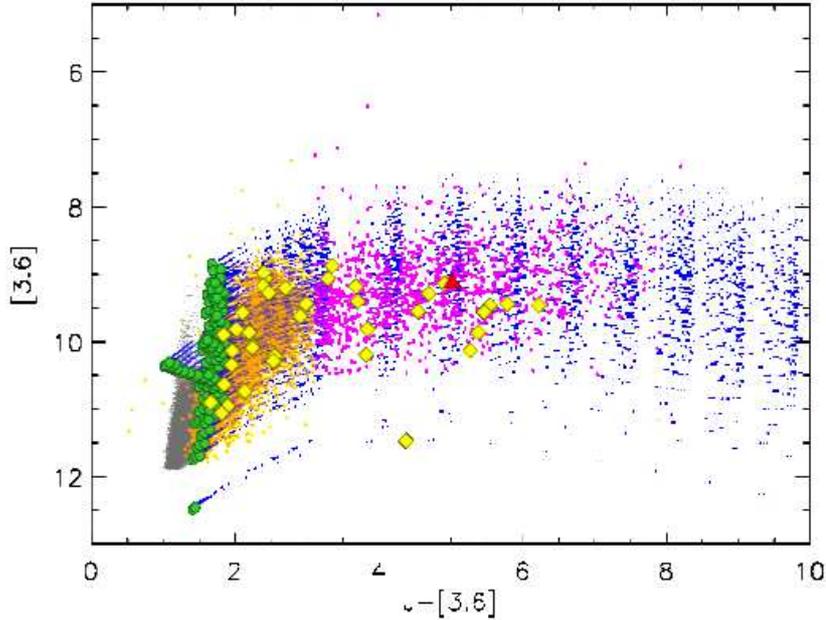}}
\caption[]{The [3.6] {\it vs.} J--[3.6] CMD, same symbols as in Fig. \ref{fig:jkcmd}. The cloud of models with $\tau_{11.3}=0.3$ is seen at J--[3.6] $\sim$ 5 mag.\label{fig:jlcmd}}
\end{figure*}

This CMD is used to distinguish between stars with oxygen-rich and carbon-rich dust chemistries (see Paper~I and references therein) because of a clear division of the stellar distribution into two branches -- the bluer branch, consisting of the O--rich AGB candidates, extends to brighter K$_{\rm s}$\ magnitudes where the red supergiants are located. The red branch reaches a maximum brightness with increasing color, at which point the near-infrared flux becomes progressively attenuated by the increasing amount of circumstellar dust. This branch consists of the C--rich and extreme AGB candidates. The carbon stars along the boundary between the two branches ({\it i.e.}, the bluest carbon stars) contain little or no dust around them; the GRAMS\ models for low optical depths (as well as the bare \citet{Aringeretal2009} photospheres) should lie along this boundary. Fig. \ref{fig:jkcmd} shows the K$_{\rm s}$\ {\it vs.} J--K$_{\rm s}$\ CMD for SAGE AGB candidates, with the model grid points superimposed. As expected, most of the A09 photospheres lie along the O/C boundary on this diagram. The NIR fluxes of AGB stars are significantly affected by pulsation; for instance, the data of \citet{Whitelocketal2003} for selected LMC carbon stars suggests a median peak-to-peak variability of $\sim$1.0 mag in the J and K$_{\rm s}$\ bands. The agreement between the data and models is very good, considering that the A09 models do not include the effects of dynamical processes. A small number of A09 models possess much bluer colors and run across the O--rich AGB population. These are the warmer ($T_{\rm eff}$\ $>$ 3500 K) photospheres which, as mentioned in Sect. \ref{subsec:photmodels}, are probably more representative of the post-AGB phase. Overall, the GRAMS\ grid provides excellent coverage of the J--K$_{\rm s}$\ colors observed for C--rich candidates and the extreme AGB candidates with detections in the near-infrared as well as carbon stars identified in the SAGE-Spec program. Photometry for OGLE LMC LPV 28579, which was studied in Paper~III, is also shown. A series of vertical bands are visible in the models; these bands correspond to models with similar optical depths. For instance, the band with $\tau_{11.3}$ = 0.3 is seen just blueward of the photometry for OGLE LMC LPV 28579.

\citet{Aringeretal2009} noted that the H--K$_{\rm s}$\ colors predicted by their models were systematically bluer than observations. They explained the disagreement at warmer temperatures as probably arising from the scaling of their C$_2$ opacity. We observe a similar disagreement between our models and SAGE data in the H band. For this reason, we do not discuss the 2MASS color--color diagram or CMDs involving the H band.

\subsubsection{[3.6] {\it vs.} J--[3.6] CMD}
This CMD, shown in Fig. \ref{fig:jlcmd}, is used in \citet{Blumetal2006} and Paper~I to select extreme AGB stars. It is somewhat similar to the K$_{\rm s}$\ {\it vs.} J--K$_{\rm s}$\ CMD. The 3.6 $\mu$m\ magnitude is less affected by dust reddening and is therefore a better luminosity proxy for the extreme AGB candidates than the K$_{\rm s}$\ magnitude. The GRAMS\ grid shows good coverage of the entire range of J--[3.6] color, reproducing the colors of almost all the extreme AGB stars with J-band detections. Once again, the models form vertical bands of increasing optical depth. OGLE LMC LPV 28579 is seen to lie on the $\tau_{11.3}=0.3$ group of models. However, there are many C--rich candidates with J--[3.6] colors up to $\sim$ 0.3 mag bluer than the model photospheres. This discrepancy in color can arise from misclassification of O--rich AGB stars as carbon-rich (resulting from our somewhat artificial definition of the O/C boundary based on near-IR colors) as well as the photometric uncertainty associated with the SAGE 3.6 $\mu$m\ photometry for our list of AGB candidates, which can be as high as 0.2--0.3 mag\footnote{Also see the SAGE Data Products Description file available at {\tt http://irsa.ipac.caltech.edu/data/SPITZER/SAGE/doc/\\SAGEDataProductsDescription$\_$Sep09.pdf}}.

\begin{figure*}[!htb]
\center{\includegraphics[width=12cm]{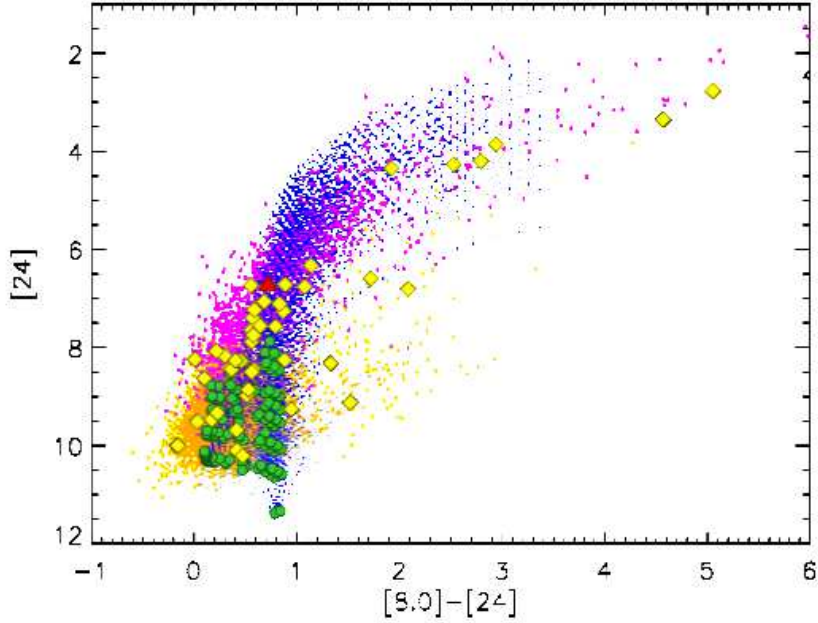}}
\caption[]{The [24] {\it vs.} [8.0]--[24] CMD, same symbols as in Fig. \ref{fig:jkcmd}.\label{fig:opcmd}}
\end{figure*}

\subsubsection{[24] {\it vs.} [8.0]--[24] CMD}
Fig. \ref{fig:opcmd} shows the [24] {\it vs.} [8.0]--[24] CMD. In Paper~I, we used this diagram to illustrate the ``bright" and ``faint" O--rich AGB populations observed in SAGE data. The carbon stars are superimposed over most of the bright O--rich population, while the faint population is clearly seen at [8.0]--[24] $\raisebox{-.4ex}{\rlap{$\sim$}} \raisebox{.4 ex}{$>$}$ 1 mag and [24] $\raisebox{-.4ex}{\rlap{$\sim$}} \raisebox{.4 ex}{$>$}$ 8 mag. The GRAMS\ models cover the entire range of MIPS 24 $\mu$m\ magnitudes observed for LMC carbon-stars. We note that there are several SAGE C--rich candidates as well as some SAGE-Spec carbon stars with [8.0]--[24] colors up to $\sim$0.5 mag bluer than the bluest GRAMS\ models (in fact, bluer than the A09 photospheres). While this discrepancy might potentially hint at limitations of the underlying stellar models used to generate the grid, it is not inconsistent with the spread expected due to photometric uncertainties and variability. Using the uncertainties from Paper~I for our C--rich and extreme AGB candidates, we estimate that the 3$\sigma$ uncertainty in the [8.0]--[24] color can be up to about 0.55 mag (For more details, we refer the reader to the SAGE Data Products Description file). We calculate the median change in the color between the two epochs of SAGE observations for the variable stars from \citet{Vijhetal2009} to be around 0.1 mag. The observed discrepancy could thus be a combined effect of these uncertainties. 

There are also about 40 C--rich and extreme AGB candidates with [8]--[24] colors redder than the reddest models available ($\sim$3.3 mag). We note here that at redder colors the extreme AGB candidate sample may be contaminated by young stellar objects (see discussion in Paper~I). However, three of the very red sources mentioned above -- IRAS~04535--6616, 2MASS~J05031662--6549450 and OGLE~LMC~LPV~16169 (not seen in Fig. \ref{fig:opcmd} due to its extremely red color of 6.02 mag) -- have been identified as carbon stars from their SAGE-Spec spectra. The first two are known C--rich AGB candidates; \citet{vanLoonetal1997} discovered the near-IR counterpart of IRAS~04535--6616 and classified it as a carbon star based on its colors, while \citet{Kontziasetal2001} identified the carbonaceous nature of 2MASS~J05031662--6549450 based on the detection of Swan C$_2$ bands in its optical spectrum. OGLE~LMC~LPV~16169 is seen projected against NGC~1835, a low-metallicity ([Fe/H] = --1.8) globular cluster with a considerable number of RR Lyrae variables \citep[{\it e.g.},][]{Walker1993}. As an extremely dust-enshrouded star, its presence in the metal-poor, low-mass population of NGC~1835 makes OGLE~LMC~LPV~16169 a very interesting object worth studying in detail. To be able to model such extremely reddened sources, we will include higher optical depth models in future versions of the grid, and also consider more dust species.

\subsubsection{K$_{\rm s}$--[3.6] {\it vs.} J--K$_{\rm s}$\ CCD}

\begin{figure*}[!htb]
\center{\includegraphics[width=12cm]{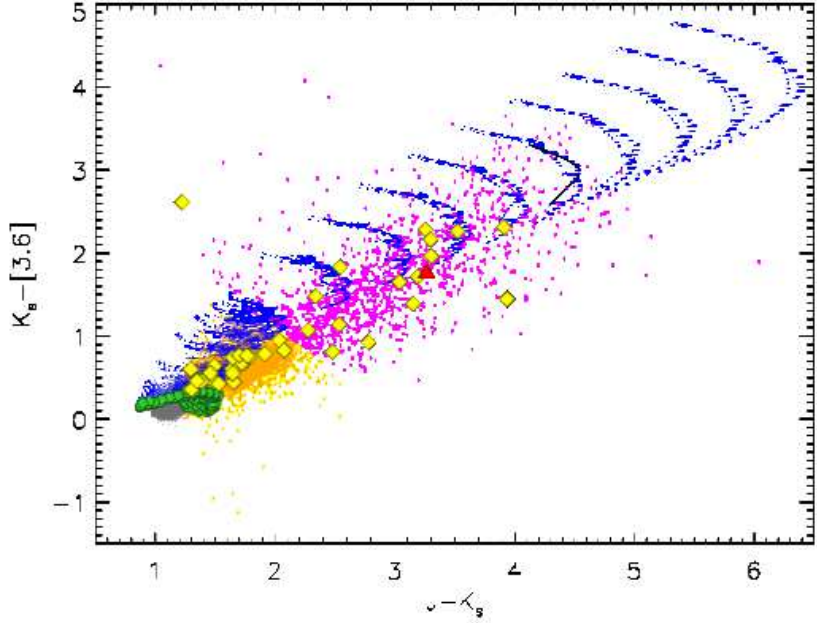}}
\caption[]{The K$_{\rm s}$--[3.6] {\it vs.} J--K$_{\rm s}$\ CCD, same symbols as in Fig. \ref{fig:jkcmd}. The ``arc" slightly blueward of OGLE LMC LPV 28579 consists of the $\tau_{11.3}=0.3$ models. The variation of the J--K$_{\rm s}$\ color with effective temperature of the central star is depicted by the solid black curve for $\log{g}$ = 0, $M$ = 2 M$_\odot$, C/O = 2, $\tau_{11.3}$ =  0.6 and $R_{\rm in}$\ = 7 $R_{\rm star}$ ($T_{\rm eff}$\ decreases in the direction of increasing K$_{\rm s}$--[3.6] color).\label{fig:jkklccd}}
\end{figure*}

\begin{figure*}[!htb]
\center{\includegraphics[width=12cm]{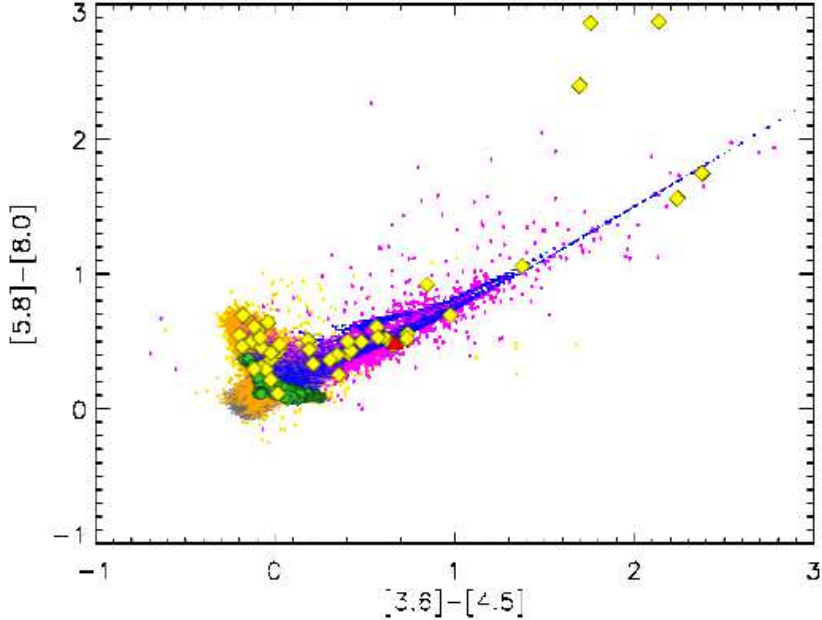}}
\caption[]{The [3.6]--[4.5] {\it vs.} [5.8]--[8.0] CCD, same symbols as in Fig. \ref{fig:jkcmd}.\label{fig:lmnoccd}}
\end{figure*}

Fig. \ref{fig:jkklccd} shows the K$_{\rm s}$--[3.6] {\it vs} J--K$_{\rm s}$\ two-color diagram, which compares the emission from the warmest regions of the dust shell to that from the stellar photosphere. We show the GRAMS\ carbon-star grid superimposed on the SAGE AGB candidates from Paper~I and the SAGE-Spec carbon stars on this diagram. The GRAMS\ models separate into groups of equal optical depth (in Fig. \ref{fig:jkklccd}, OGLE LMC LPV 28579 is seen to fall close to the group of models with optical depth $\tau_{11.3}$ = 0.3). This is more obvious at redder colors because of the coarse sampling at higher optical depths in the current version of the model grid. The J--K$_{\rm s}$\ color shows an interesting trend with effective temperature of the central star: for each set of models with the same optical depth and inner shell radius, decreasing $T_{\rm eff}$\ first causes a reddening in J--K$_{\rm s}$, due to the increasing continuum flux in the K$_{\rm s}$\ band relative to the J band. However, a lower $T_{\rm eff}$\ value also means increased strength in the $\sim$2.3 $\mu$m\ CO absorption feature, which eventually counteracts the increase in continuum K$_{\rm s}$\ flux and shifts the coolest models towards bluer J--K$_{\rm s}$. This trend with $T_{\rm eff}$\ is depicted in Fig. \ref{fig:jkklccd} for the GRAMS\ models with $\log{g}$ = 0, $M$ = 2 M$_\odot$, C/O = 2, $\tau_{11.3}$ =  0.6 and $R_{\rm in}$\ = 7 $R_{\rm star}$. At low optical depths ($\tau < 0.1$), the K$_{\rm s}$--[3.6] color is affected by the K$_{\rm s}$\ band CO absorption feature as well as strong HCN + C$_2$H$_2$ absorption near 3 $\mu$m. With decreasing $T_{\rm eff}$, the K$_{\rm s}$--[3.6] color first moves to bluer values, reaches a minimum then reverses this trend. This turnover trend is not seen in either color for GRAMS\ models with the highest optical depths ($\tau > 1$) due to the large amounts of dust which overwhelms emission from the central star. For these models, a decrease in effective temperature is accompanied by a reddening in both colors.

The GRAMS\ models are able to reproduce the range of observed colors in this diagram. The colors of the bare photospheres are similar to the bluest carbon-star candidates on the boundary with the O--rich candidates, as would be expected if these represented carbon stars with little or no dust. However, at redder J--K$_{\rm s}$\ colors, a significant fraction of C--rich and extreme AGB candidates are not covered by the models; in fact these sources are up to 0.8 mag bluer in K$_{\rm s}$--[3.6] than the models. We have already mentioned that the photometric uncertainty in the 3.6 $\mu$m\ SAGE data can be up to a few tenths of a magnitude. This uncertainty alone is not sufficient to explain the observed discrepancy. LMC carbon stars are known to exhibit strong C$_2$H$_2$ absorption in the 3 $\mu$m\ region; as discussed previously, this can depress the flux in the 3.6 $\mu$m\ band. At higher dust optical depths, emission from the circumstellar dust may somewhat reduce this effect \citep[see, {\it e.g.},][]{vanLoonetal2006}, which would explain why our high $\tau_{11.3}$ models reproduce the colors of the reddest extreme AGB candidates on this plot.

\subsubsection{[5.8]--[8.0] {\it vs.} [3.6]--[4.5] CCD}
Fig. \ref{fig:lmnoccd} shows the IRAC-only two-color diagram. An increase in [3.6]--[4.5] color is accompanied by an increase in the [5.8]--[8.0] color for the extreme AGB candidates. This is simply a result of increasing dust emission in the mid-IR due to increasing optical depth. A significant fraction of the C--rich AGB candidates, however, show a decreasing [5.8]--[8.0] color with redder [3.6]--[4.5] colors. This trend is also seen to a smaller extent in the distribution of the bluest ([3.6]--[4.5] $<$ 0) model grid points. There is a strong CO absorption feature in the 5.8 $\mu$m\ band (see, {\it e.g.}, Fig. 2 in Paper~III) which could be filled in by emission from dust with increasing optical depth.

The model colors agree well with those observed for the data, especially for the extreme AGB candidates. The photometric uncertainties as well as pulsation amplitudes in the IRAC bands are lower than in the near-IR, which may explain the smaller spread in the data around the models. At the bluest colors, there are many C--rich AGB candidates that are bluer than the model photospheres. This could be explained, once again, on the basis of the absorption features in the IRAC 3.6 (HCN + C$_2$H$_2$) and 5.8 $\mu$m\ (CO) bands.

\begin{figure*}[!hpbt]
\begin{center}$
\begin{array}{cc}
\includegraphics[width=6cm]{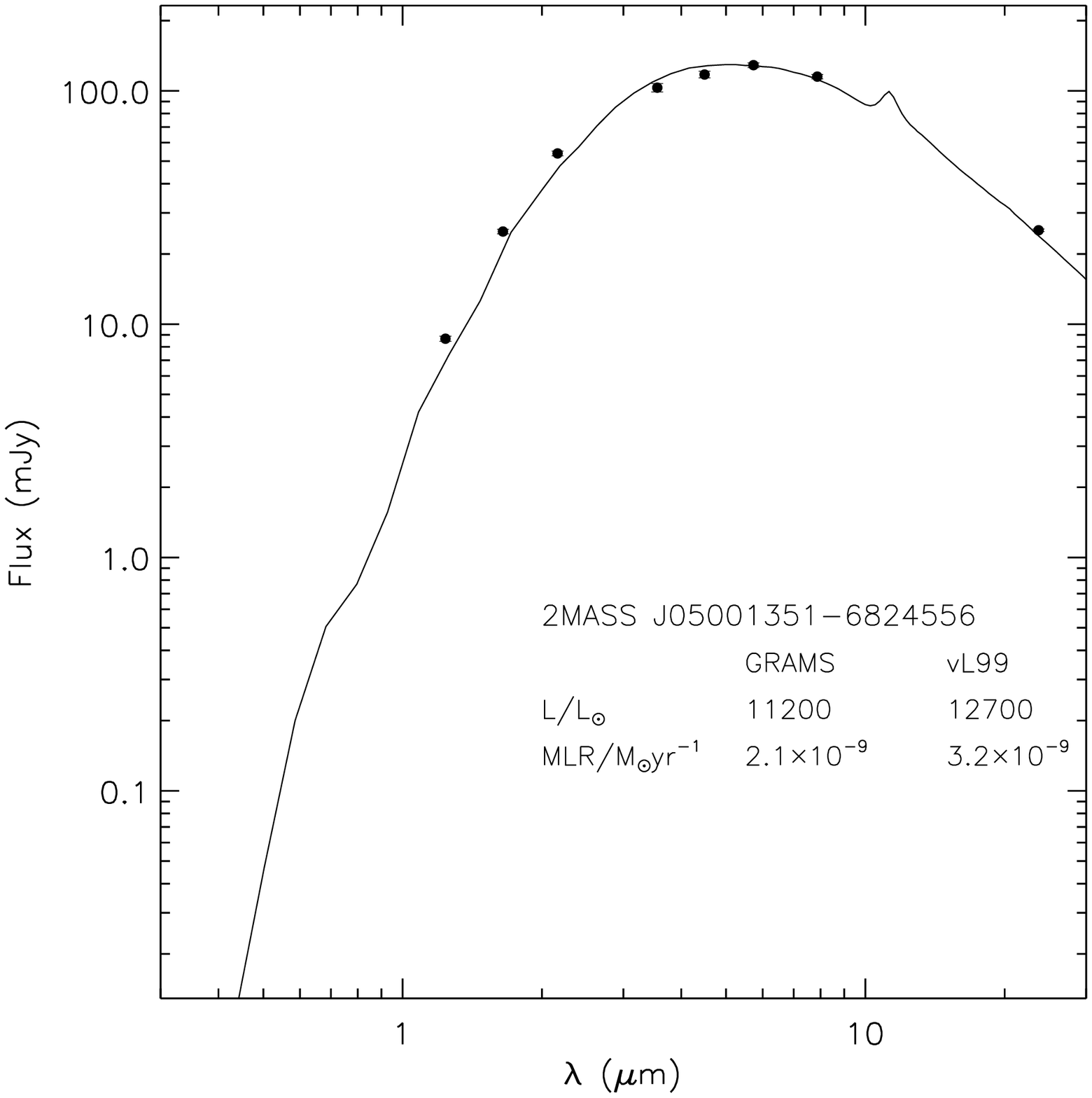} &
\includegraphics[width=6cm]{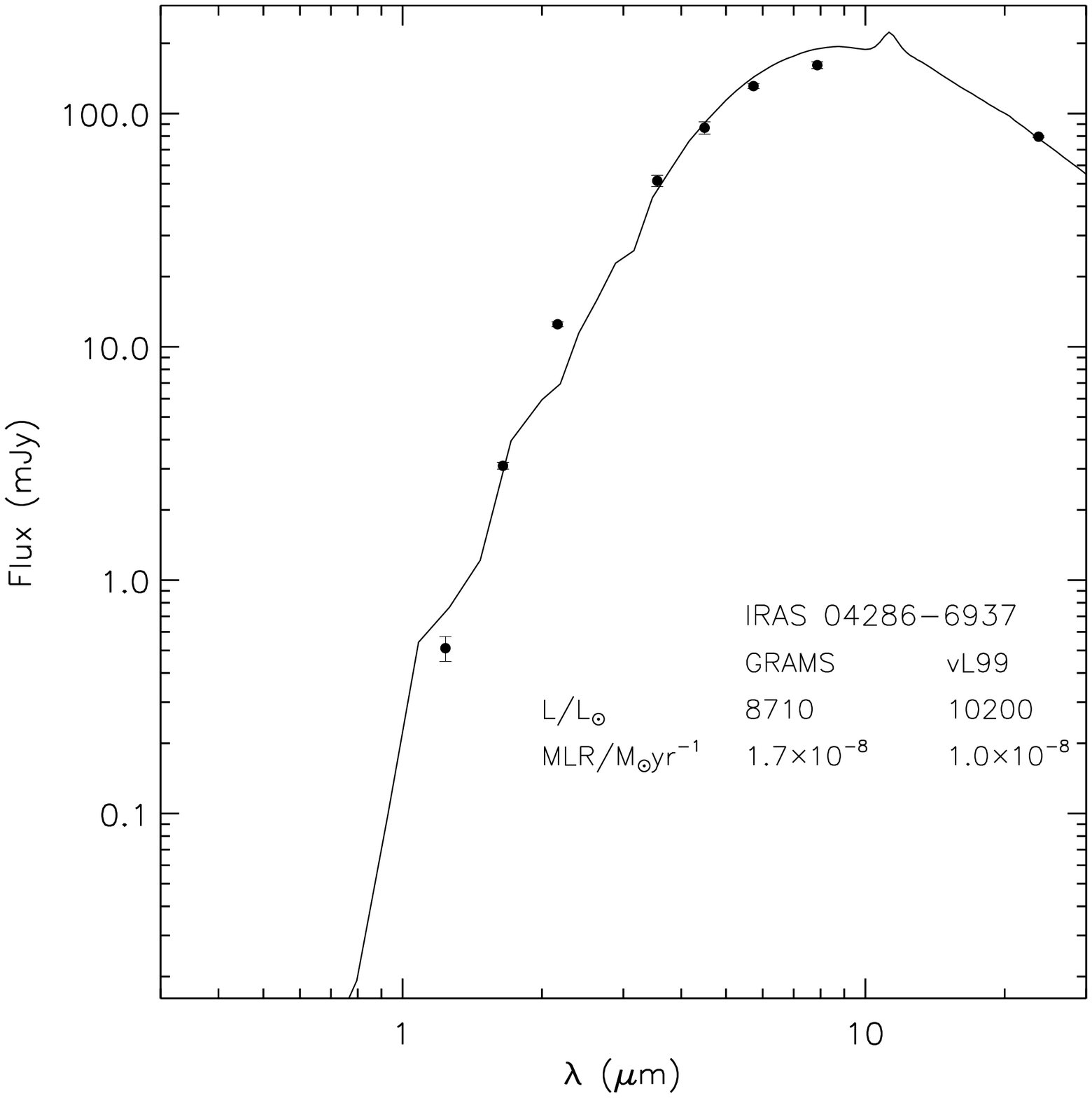} \\ \includegraphics[width=6cm]{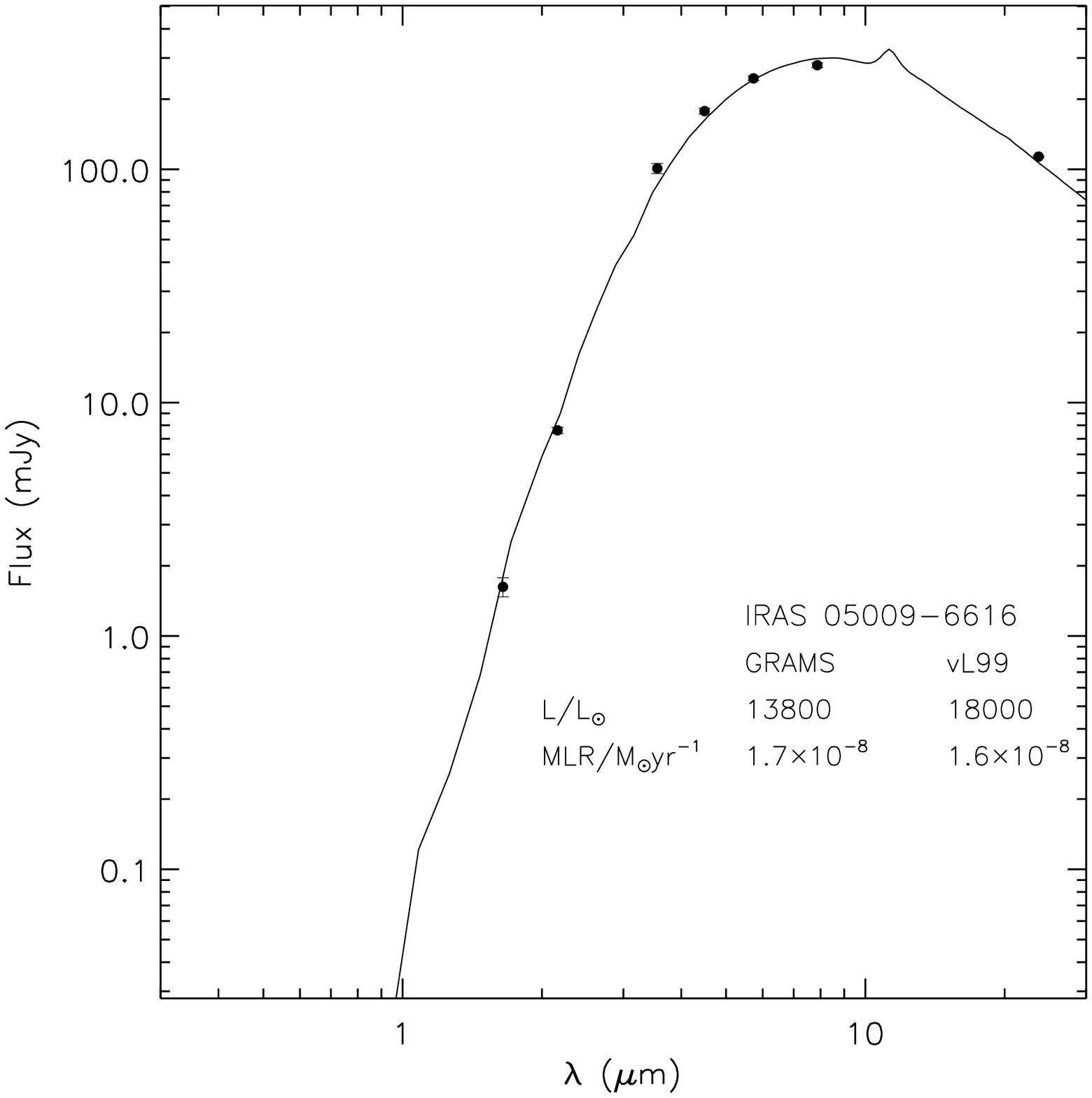} &
\includegraphics[width=6cm]{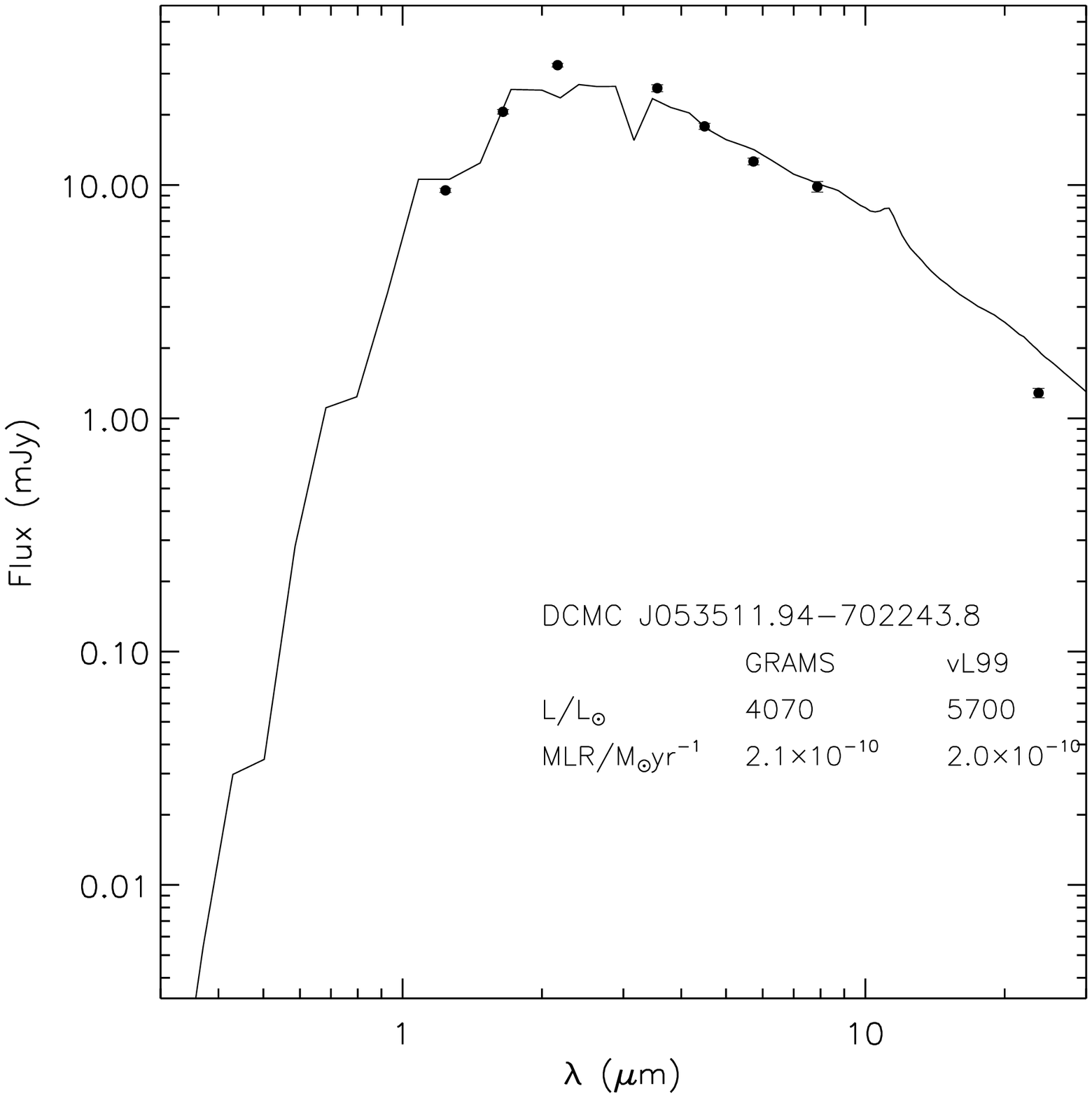}
\end{array}$
\end{center}
\caption{GRAMS\ fits to the SEDs of four carbon stars modeled by \citet{vanLoonetal1999}. \label{fig:vL99fits}}
\end{figure*}

We also note that the carbon-star models have continuous coverage of the extreme AGB candidates over the entire range of colors. Compare this to the oxygen-rich models from Paper~IV, which do not reproduce the colors of the stars in the range 0.2 $<$ [3.6]--[4.5] $<$ 1 mag (see Fig. 5 and the discussion in Sect. 4.1.2 in Paper~IV). This difference in relative coverage can be used to identify the chemical types of the extreme AGB candidates through SED fitting.

\section{Discussion}
\subsection{Preliminary SED fits}
\subsubsection{Fits to the \citet{vanLoonetal1999} sources}
\label{subsec:vL99comparison}

We are ultimately interested in fitting the SEDs of the entire SAGE AGB star candidate list. In this section, we evaluate simple chi-squared fits to the SEDs of spectroscopically confirmed carbon stars in the LMC for which detailed models already exist in the literature. Specifically, we consider the sources studied by \citet{vanLoonetal1999} and \citet{Groenewegenetal2009} (hereafter, vL99 and G09 respectively). We also compare the results of such SED fitting of the carbon star OGLE LMC LPV 28579 with the parameters derived from detailed modeling of this source in Paper~III. The chi-squared fitting considered in this section allows a simple consistency check of the GRAMS\ grid. This technique is able to predict accurate luminosities and dust mass-loss rates (see Paper~III), but it may not be able to produce strong constraints on other parameters. For instance, we expect that a set of reasonable fits to a reddened source ({\it e.g.}, a carbon star from the vL99 or G09 sample) will exhibit considerable degeneracy in predicted stellar parameters. 
In what follows, therefore, we only compare the luminosities and dust mass-loss rates for the vL99 and G09 sources with the corresponding GRAMS\ best-fit values. We use the gas:dust ratio provided in vL99 and G09 (500 and 200 respectively) to convert their total mass-loss rates to dust rates. Our $\upsilon_{\rm exp}$ value is kept fixed at the same value as that of G09. However, vL99 use a luminosity-dependent expansion velocity given by
\begin{eqnarray}
\label{vexpscaling}
{\upsilon_{\rm exp}\over 10~{\rm km~s}^{-1}} = \left({L\over 3\times 10^4~{\rm L}_\odot}\right)^{1/4}
\end{eqnarray}
For each vL99 source, we compare both the ``scaled" ($\upsilon_{\rm exp}$ as in Eq. \ref{vexpscaling}) and ``unscaled" ($\upsilon_{\rm exp} = 10~{\rm km~s}^{-1}$) versions of the dust mass-loss rate with our best-fit value in Sect. \ref{subsec:vL99comparison}.

Evolved AGB stars have significant variability amplitudes at visible wavelengths; this variation must be taken into account in order to reconstruct the SED from multi-wavelength photometry obtained at different pulsation phases. We will consider these details in future papers. Presently we fit only the JHK$_{\rm s}$ and {\it Spitzer} bands for these sources. As discussed in Sect. \ref{subsec:results:lummlrcoverage}, our models provide substantial coverage of the range of observed luminosities. Moreover, at the moment we would like to demonstrate the general agreement between fits from our model grid and those from detailed studies. For these reasons, we do not vary the model luminosities in order to find the best fit. We will treat the luminosity scale as an extra free parameter when we consider SED fitting in detail.

\subsubsection{Fits to the \citet{Groenewegenetal2009} sources}
\label{subsubsec:G09compare}
\begin{figure*}[!hptb]
\begin{center}$
\begin{array}{cc}
\includegraphics[width=6cm]{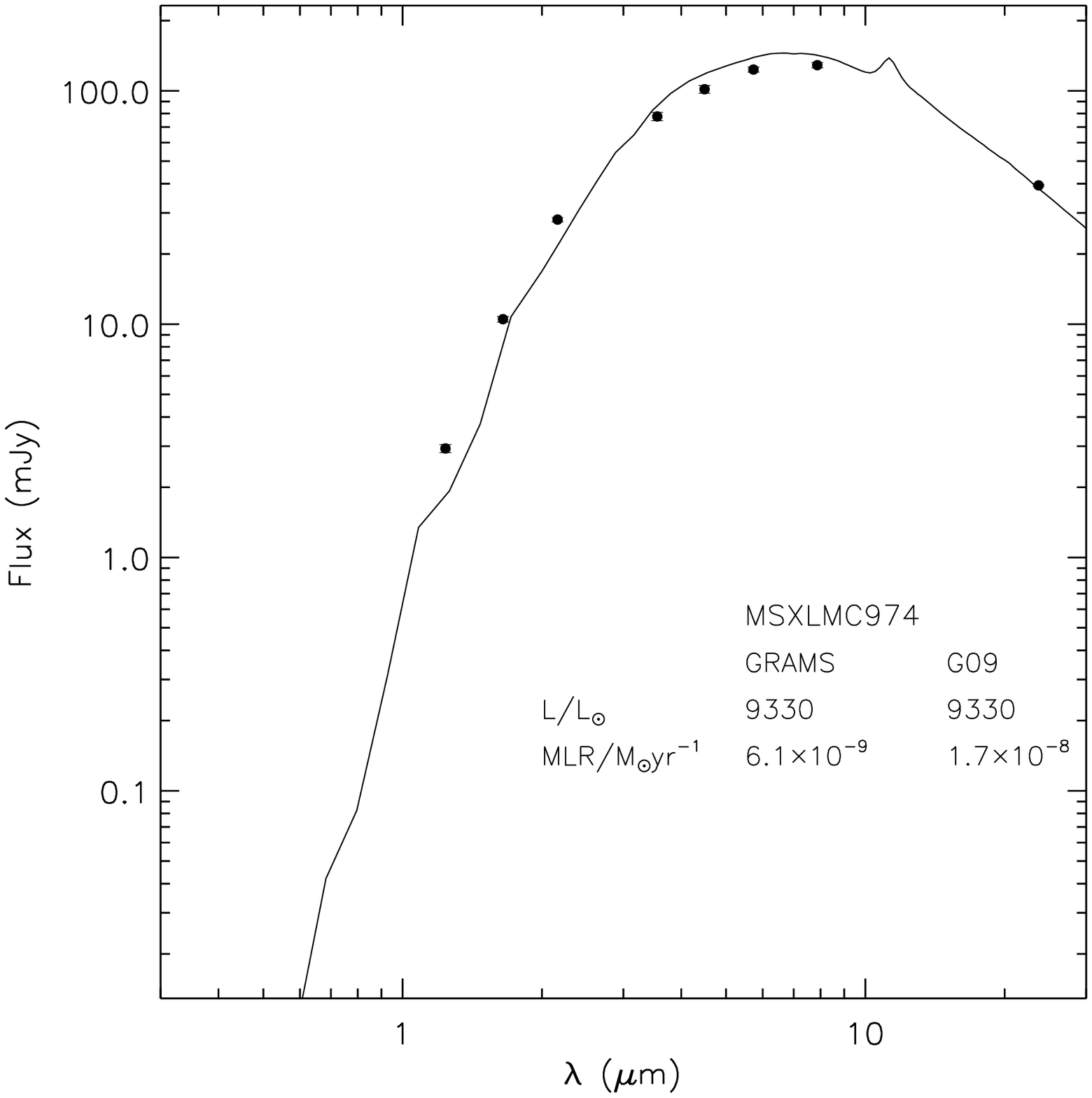} &
\includegraphics[width=6cm]{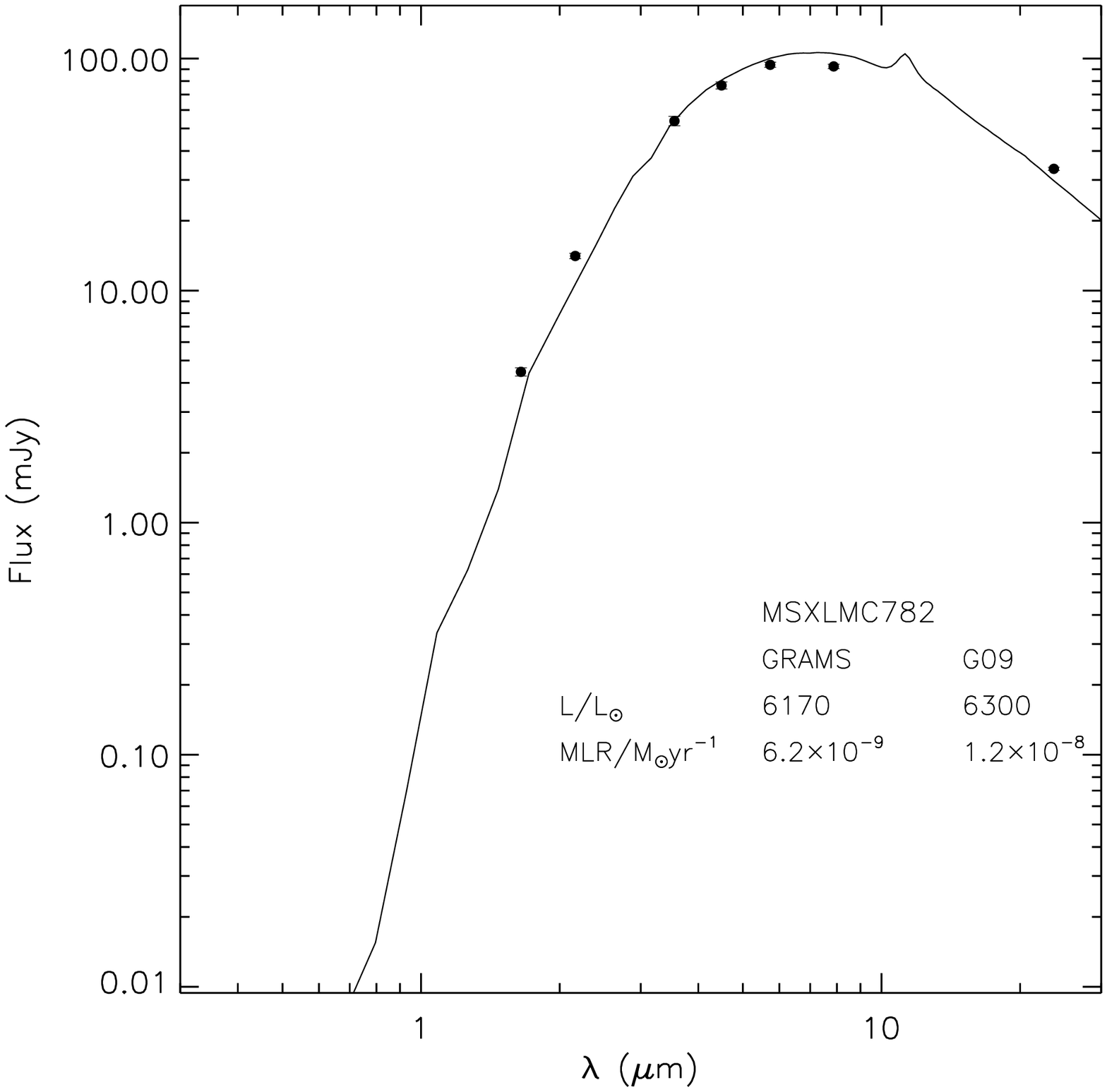} \\ \includegraphics[width=6cm]{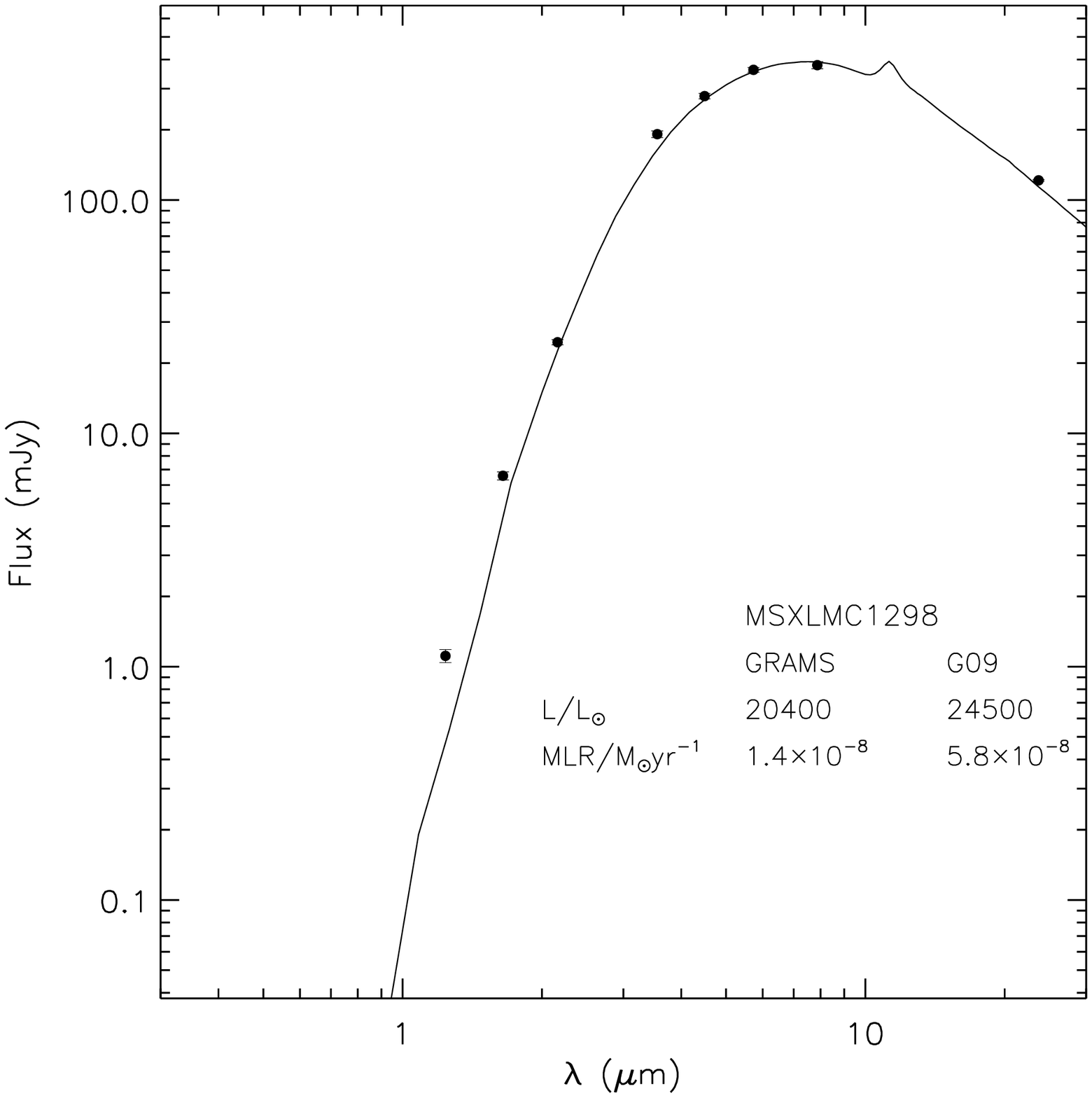} &
\includegraphics[width=6cm]{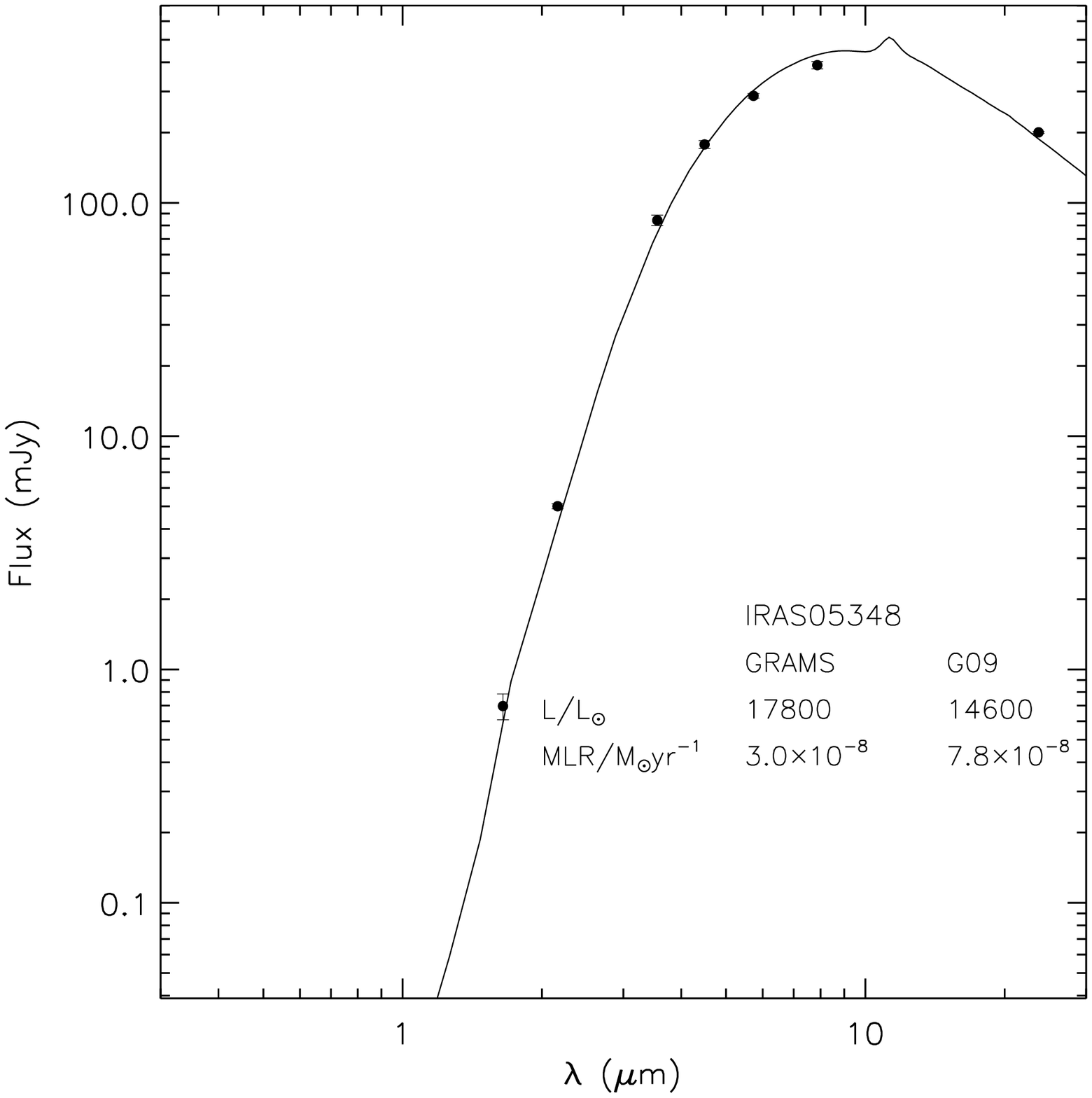}
\end{array}$
\end{center}
\caption{GRAMS\ fits to the SEDs of four carbon stars modeled by \citet{Groenewegenetal2009}. \label{fig:G09fits}}
\end{figure*}

\begin{figure}[!hptb]
\resizebox{\hsize}{!}{\includegraphics{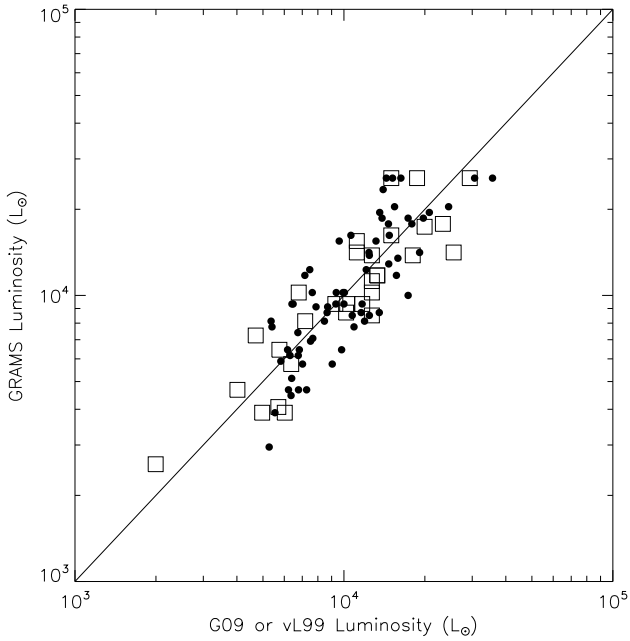}}
\caption{Luminosities calculated from GRAMS\ SED fitting plotted against the luminosities obtained from detailed modeling by  \citet{vanLoonetal1999} (squares) and \citet{Groenewegenetal2009} (circles). The solid line represents a 1:1 agreement. \label{fig:comparelum}}
\end{figure}

\begin{figure}[!htpb]
\resizebox{\hsize}{!}{\includegraphics{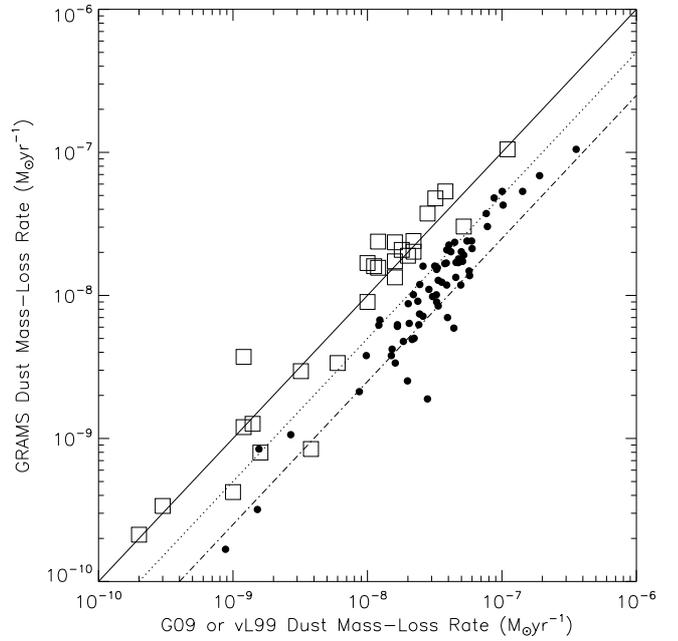}}
\caption{Same as Fig. \ref{fig:comparelum}, but for dust mass-loss rates. Also shown are lines along which the GRAMS\ mass-loss rate is lower than the rates from detailed modeling (vL99 or G09) by a factor of 2 and 4 (dotted and dashed line respectively).\label{fig:comparemlr}}
\end{figure}

In their paper, van Loon et al. modeled 31 carbon stars (57 LMC evolved stars total) with ISO photometry supplemented by ground-based near-IR photometry. ISO spectroscopy was also available for 10 of these stars. In Paper~I, we used the vL99 mass-loss rates  to estimate the total dust injection rate into the LMC from the entire AGB population. We found nearest-neighbor matches in our AGB candidate list (see Paper I) for all of the van Loon et al. carbon stars\footnote{We excluded IRAS 05289--6617 from further analysis because it lacked detections in the optical and near-IR bands as well as most of the {\it Spitzer} bands \citep[see discussion in][]{vanLoonetal1999}.}. We fit GRAMS\ models to each of these stars, fitting only the near- and mid-IR bands as previously mentioned. Fig. \ref{fig:vL99fits} shows best-fit GRAMS\ spectra for four of the vL99 stars. Overall, our best-fit spectra reproduce the observed photometry well. The best-fit luminosities and dust mass-loss rates also show a good agreement with the values calculated by van Loon et al., as shown in Figs. \ref{fig:comparelum} and \ref{fig:comparemlr}. These plots demonstrate the power of the simple model grid fitting technique to obtain reliable luminosities and mass-loss rates. The GRAMS\ luminosities are within a factor of $\sim$ 1.75 of the vL99 values. The median discrepancy is --13\%, with a standard deviation of 29\% around this value. With the exception of TRM 45, for which we obtain a poor fit, our mass-loss rates agree with those of vL99 (unscaled) to within a factor of 4.5. Our rates are discrepant from the unscaled vL99 rates by a median value of 8\% with a standard deviation of 83\% and from the scaled versions by --13\% with a standard deviation of 125\%.

Groenewegen et al. used {\it Spitzer} IRS spectra to model 101 Magellanic Cloud carbon stars, using multi-epoch photometry to constrain the optical variability. Their dataset included 68 LMC carbon stars. We found counterparts in our list of AGB candidates for all but three of the G09 stars -- MSX LMC 95, MSX LMC 1384 and NGC 1978 IR1. However, on searching the SAGE Archive point source lists, we were able to find a match for MSX LMC 95 in the IRAC Epoch 1 Archive. We also found detections for MSX LMC 1384 and NGC 1978 IR1 in the IRAC Mosaic Photometry catalog. Fig. \ref{fig:G09fits} shows the best-fit GRAMS\ spectra for four carbon stars studied by \citet{Groenewegenetal2009}. The best-fit luminosities and dust mass-loss rates for the entire set are compared with the G09 estimates in Figs. \ref{fig:comparelum} and \ref{fig:comparemlr} respectively. Once again, the luminosities predicted by GRAMS\ SED fitting are within a factor of $\sim$ 1.75 of the values determined from detailed modeling. The median discrepancy is very low at --1\%, with a standard deviation of 31\%. However, the behavior of the mass-loss rates is different from that observed with the vL99 sample. 

While the luminosities are more or less evenly distributed about the 1:1 line for GRAMS\ fits to both the vL99 and G09 sources, our dust mass-loss rate estimates for most of the G09 sources are systematically lower than theirs by a factor of about 2--4. The median discrepancy in this case is --168\%, with a standard deviation of almost 200\% around this value. This discrepancy is due to our different choice of amorphous carbon optical constants. The \citet{RouleauMartin1991} optical constants used in G09 give rise to a consistently lower absorption efficiency over the $\sim$0.1--100 $\mu$m\ range \citep[see, {\it e.g.}, Fig. 1 of][]{Suh2000}, thus requiring a higher dust shell mass in order to fit a given mid-IR flux. This in turn increases the calculated mass-loss rate. In order to verify the effect of this change, we performed detailed radiative transfer calculations on ten of the G09 sources using the GRAMS\ best-fit parameters from SED fitting as input, with two modifications: we used the \citet{RouleauMartin1991} optical constants for amorphous carbon and set the SiC mass fraction to the value determined in G09 for each star. We found that in general the \citet{RouleauMartin1991} optical constants gave mass-loss rates that were 2--4 times higher than those predicted by the GRAMS\ best fits, explaining the systematic offset observed in Fig. \ref{fig:comparemlr}. For comparison, the \citet{Preibischetal1993} amorphous carbon optical constants used by \citet{vanLoonetal1999} have absorption efficiencies similar to that of the \citet{Zubkoetal1996} dust model. We computed {\bf 2D}ust models for five of the vL99 sources with the \citet{Preibischetal1993} dust constants and found that the resulting dust mass-loss rates were at most 12\% different from those predicted by the corresponding GRAMS best-fit models. In developing this first version of the model grid, we used a single set of dust properties. We will investigate the effect of varying the dust properties in detail in future versions of the grid.

\begin{figure}[!htb]
\resizebox{\hsize}{!}{\includegraphics{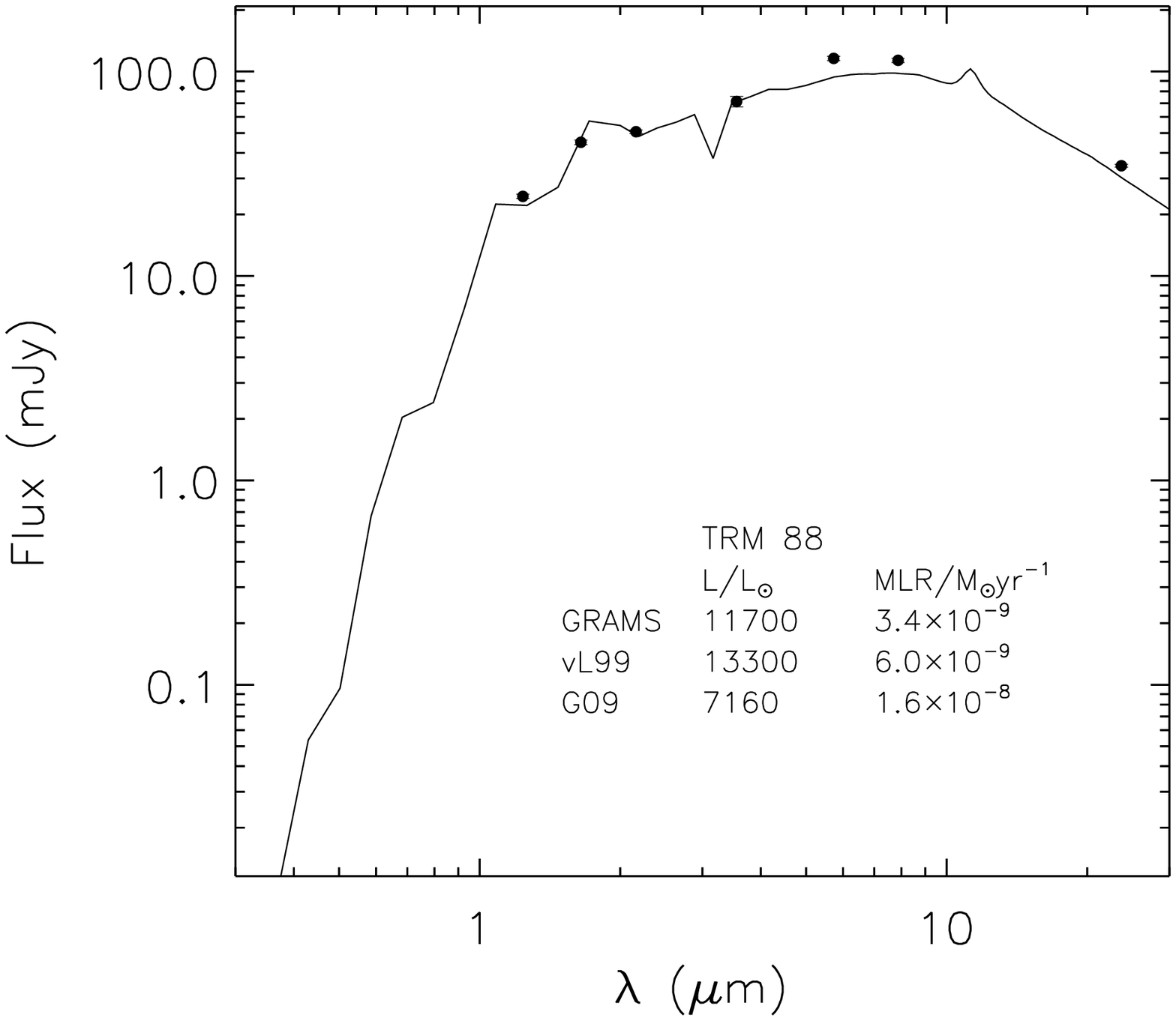}}
\caption{GRAMS best-fit for TRM 88. The best-fit luminosity and mass-loss rates are more consistent with the vL99 estimates.\label{fig:compareTRM88}}
\end{figure}

TRM 88 was modeled by both vL99 and G09. Fig. \ref{fig:compareTRM88} shows the GRAMS\ best-fit spectrum to the observed photometry of TRM 88. The GRAMS\ best-fit luminosity (11\,700 L$_\odot$) and mass-loss rate ($3.4\times 10^{-9}$ M$_\odot$ yr$^{-1}$) are closer to the values estimated by van Loon et al. In this case, the GRAMS\ fit is somewhat poor as it under-predicts the mid-IR emission resulting in a low mass-loss rate. The marginal quality of the fit is partly due to the fact that we have not taken the near-IR variability into account; the fluxes are currently weighted by their photometric errors which are quite small. The best-fit model has an optical depth $\tau_{11.3}$ = 0.1 and $R_{\rm in}$\ = 12 $R_{\rm star}$. We do not have enough resolution in optical depth (the closest higher optical depth is 0.2) and inner radius (we only explore four values) to provide a better fit to the data. Despite these details, we find good agreement with the vL99 results.

\subsubsection{Fit to OGLE LMC LPV 28579}
\begin{figure}[!htb]
\resizebox{\hsize}{!}{\includegraphics{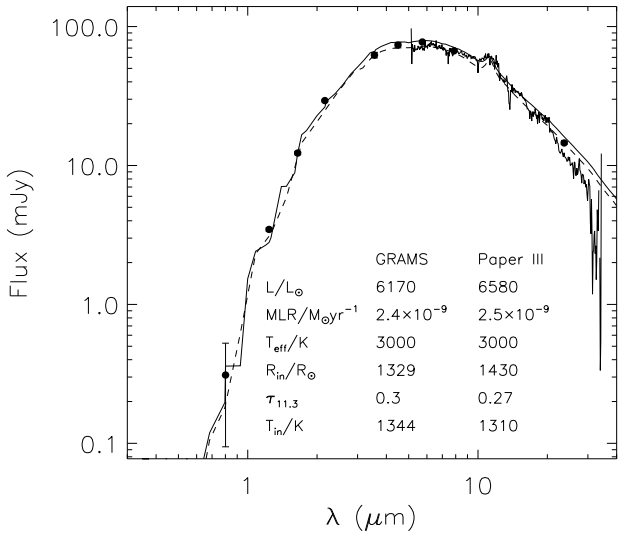}}
\caption[]{Best fit SEDs to the SAGE photometry (dots) and SAGE-Spec IRS spectrum for OGLE LMC LPV 28579. The solid curve is a result of the detailed model from Paper~III, while the dashed curve is the  GRAMS best-fit. A comparison of the fit parameters is also shown.\label{fig:oglefitcompare}}
\end{figure}

In Paper~III we presented a {\bf 2D}ust\ model best-fit model for OGLE LMC LPV 28579. The dust properties determined for this best-fit model were subsequently used for the entire carbon-star grid. Our detailed model suggested a luminosity of 6580 L$_\odot$\ for the SAGE Epoch 1 observations, an optical depth of $\tau_{11.3}=0.27$, and a dust mass-loss rate of $2.5 \times 10^{-9}$ M$_\odot$ yr$^{-1}$. The mass-loss rate derived for OGLE LMC LPV 28579 was found to be consistent with the rates derived from various empirical relationships (see discussion in Paper~III for details). In this paper we fit the observed SED for OGLE LMC LPV 28579 using models from the GRAMS\ grid. This serves as a consistency check for the model grid. The best-fit GRAMS\ SED gives a luminosity of 6170 L$_\odot$, $\tau_{11.3}=0.3$ and a dust mass-loss rate of $2.4\times 10^{-9}$ M$_\odot$ yr$^{-1}$. These numbers are within a few percent of the values calculated in Paper~III. Fig. \ref{fig:oglefitcompare} compares the best-fit models from both papers with the SAGE Epoch 1 photometry and SAGE-Spec spectrum for OGLE LMC LPV 28579. We also obtain good agreement with the values obtained for the optical depth, dust shell inner radius as well as temperature at the inner radius.

\section{Summary}
\label{sec:summary}
We constructed a grid of carbon star models using the radiative transfer code {\bf 2D}ust\ and for a range of various stellar and dust shell parameters. We intend to use these models in conjunction with our models for O--rich AGB and red supergiant stars described in \citet{Sargentetal2011} to investigate the mass-loss return from evolved stars to the LMC. The models can also be used for similar estimates from large photometric samples of C--rich AGB stars. The grid covers luminosities from 2000 L$_\odot$\ to 26\,000 L$_\odot$ by using model photospheres spanning temperatures in the range 2600--4000 K. Assuming spherically symmetric dust shells, and constant expansion velocity and mass-loss rate, we perform radiative transfer using the {\bf 2D}ust\ code for 11.3 $\mu$m\ optical depths ranging from $10^{-3}$ to 4 and values for the inner radii of 1.5, 3, 4.5, 7 and 12 times the stellar radius. This results in over 12\,000 models with dust temperatures under 1800 K, with dust mass-loss rates in the range $10^{-12}-10^{-7}$ M$_\odot$ yr$^{-1}$. 

We synthesize photometry for these models in optical as well as near- and mid-infrared bands. The entire set of models, including spectra and synthetic photometry, will soon be available at the {\bf 2D}ust website at STScI. We compare the resulting colors and magnitudes with those observed for AGB candidates in the SAGE survey and confirmed AGB stars from the SAGE-Spec survey, finding good overall agreement with these data. Using a chi-squared fitting routine, we obtain best-fit spectra, luminosities and mass-loss rates for spectroscopically identified carbon stars in the \citet{vanLoonetal1999} and \citet{Groenewegenetal2009} samples. The luminosities predicted from simple fitting to the photometry are in good agreement with those determined by these detailed models. Our mass-loss rates for the van Loon et al. sample agree well with their values. However, our rates for the Groenewegen et al. sample are lower by a factor of 2--4, most likely due to a different choice of dust properties. We find excellent agreement between the results of detailed modeling of OGLE LMC LPV 28579 from Paper~III and the fitting employed in this work.

One of the aims of our study is to provide a general-use fitter that can be applied to large sets of photometric data. This will be the focus of one of our future papers. The GRAMS\ grid will enable the assessment of mass-loss return from galaxy-wide point source catalogs from projects such as SAGE. While most extreme AGB stars are probably carbon-rich, some are very bright OH/IR stars. Currently, there is no way to clearly distinguish between these sources with photometry alone. We hope that a comparison of our results with ongoing studies such as AKARI and WISE that include filters sensitive to silicate and SiC features will enable us to define the separation between C--rich and O--rich sources in this extreme regime.

\begin{acknowledgements}
We would like to thank our referee, Jacco van Loon, for his constructive comments which helped improve the quality of the paper. We also thank Angela Speck for her valuable insight on dust condensation temperatures. This work is based on observations made with the {\it Spitzer} Space Telescope, which is operated by the Jet Propulsion Laboratory, California Institute of Technology, under NASA contract 1407. We acknowledge funding from the NAG5-12595 grant and the SAGE-LMC {\it Spitzer} grant 1275598. The authors would also like to thank Bernie Shiao at STScI for his invaluable assistance with the SAGE database.
\end{acknowledgements}

\end{document}